\documentclass[traditabstract]{aa}
\usepackage{verbatim} 
\usepackage{amsmath}
\usepackage{amssymb}
\usepackage{graphicx}
\usepackage{booktabs}
\usepackage{xspace}
\usepackage{siunitx}
\usepackage{mathrsfs}
\usepackage{placeins}
\usepackage{subcaption}
\usepackage{lscape}

\makeatletter

\usepackage{txfonts}
\usepackage{enumitem}
\usepackage{natbib,twoopt}
\usepackage[breaklinks=true,colorlinks=true,linkcolor=blue,urlcolor=blue,citecolor=blue]{hyperref} 
\bibpunct{(}{)}{;}{a}{}{,}             
\makeatletter
  \newcommandtwoopt{\citeads}[3][][]{\href{http://adsabs.harvard.edu/abs/#3}%
    {\def\hyper@linkstart##1##2{}%
     \let\hyper@linkend\@empty\citealp[#1][#2]{#3}}}
  \newcommandtwoopt{\citepads}[3][][]{\href{http://adsabs.harvard.edu/abs/#3}%
    {\def\hyper@linkstart##1##2{}%
     \let\hyper@linkend\@empty\citep[#1][#2]{#3}}}
  \newcommandtwoopt{\citetads}[3][][]{\href{http://adsabs.harvard.edu/abs/#3}%
    {\def\hyper@linkstart##1##2{}%
     \let\hyper@linkend\@empty\citet[#1][#2]{#3}}}
  \newcommandtwoopt{\citeyearads}[3][][]%
    {\href{http://adsabs.harvard.edu/abs/#3}
    {\def\hyper@linkstart##1##2{}%
     \let\hyper@linkend\@empty\citeyear[#1][#2]{#3}}}
\makeatother

\newcommand{\msun}{{\rm M}_{\odot}}

\newcommand{\rsun}{{\rm R}_{\odot}}

\newcommand{\eg}{e.g.\@\xspace}
\newcommand{\cf}{c.f.\@\xspace}
\newcommand{\ie}{i.e.\@\xspace}

\titlerunning{Contact tracing of binary stars with conservative mass transfer}
\authorrunning{J.~Henneco et al.}
\makeatother
\begin{document}
\title{Binary Evolution at the Extremes of Mass-Transfer Efficiency}
\subtitle{Contact, Mergers, and Population Signatures}

\author{Jan Henneco \inst{1,2} \corrauth{jan.henneco@protonmail.com}\and
Utkarsh R. Basu \inst{1,3,4} \email{utkarsh.basu@ru.nl}\and
Fabian R. N. Schneider \inst{1,5} \email{fabian.schneider@h-its.org}}

\institute{Heidelberger Institut f\"ur Theoretische Studien, Schloss-Wolfsbrunnenweg 35, 69118 Heidelberg, Germany\label{HITS} 
\and School of Mathematics, Statistics and Physics, Newcastle University, Newcastle upon Tyne, NE1 7RU, United Kingdom\label{NU}
\and Universit\"at Heidelberg, Department of Physics and Astronomy, Im Neuenheimer Feld 226, 69120 Heidelberg, Germany\label{UNIHD}
\and Institute for Mathematics, Astrophysics and Particle Physics, Radboud University, Heyendaalseweg 135, 6525AJ Nijmegen, Netherlands\label{RU}
\and Zentrum f\"ur Astronomie, Astronomisches Rechen-Institut (ZAH/ARI), Heidelberg University,  M\"{o}nchhofstr. 12-14, 69120 Heidelberg, Germany\label{ZAHARI}}

\date{Received 2 July 2026 / Accepted 4 August 2026}

\abstract{Binary star evolution remains inherently uncertain, and several physical processes are still not well understood. These open questions in binary physics come in addition to major uncertainties in single-star evolution, such as angular momentum transport and interior mixing. For example, the efficiency of mass transfer -- which is the fraction of transferred mass that is actually accreted -- is still one of the main uncertainties in binary evolution. In this work, we present a new grid of 1D binary evolution models with identical initial conditions to an earlier grid in which mass transfer was limited by the spin-up of the accretors. In this new grid, we employ fully conservative mass transfer, allowing us to do a one-to-one comparison between the two grids, and covering the full possible range from highly non-conservative to fully conservative mass transfer. We explore how these two maximally different mass-transfer efficiencies change the occurrence and incidence of contact phases, stellar mergers and common envelope phases, and how they affect the present-day population of (post-)mass-transferring binaries. We find that fully conservative mass transfer increases the incidence of contact systems by roughly a factor of 6 (from 11\% to 62\%), and the incidence of stellar mergers by more than a factor of 2 (from 16\% to 38\%). We also find the emergence of double-core CE phases, which are absent for lower mass-transfer efficiencies. Comparing two synthetic binary populations built using the two grids with observed Algol and stripped-star binaries reveals that even though conservative mass transfer is favoured to reproduce the observed stripped-star binaries, the observed population of Algol binaries cannot be explained by a single mass-transfer efficiency. We conclude that the mass-transfer efficiency depends on the configuration of binary systems and cannot be described by a single value. Our work highlights the need for a better understanding of binary mass transfer and indicates that current binary-star models are incomplete.}


\maketitle
\nolinenumbers
\defcitealias{Henneco2024a}{Paper~I}

\section{Introduction}\label{sec:introduction}
Binary stars play a vital role in many facets of stellar astrophysics; yet, our understanding of binary evolution is still limited and significant uncertainties remain. When the initially more massive star in a binary system fills its Roche lobe, the ensuing mass-transfer phase alters the further evolution of the mass-losing donor and mass-gaining accretor star (see \citealt{MarchantBodensteiner2024} for a recent review). For example, our understanding of the stability and efficiency of these mass-transfer phases influences our predictions for, among many other things, the progenitors of gravitational wave sources \citep[\eg][]{Tauris2017,Mandel2022}, supernova progenitors \citep[\eg][]{Podsiadlowski1992a, Claeys2011, Langer2012, Schneider2021, Schneider2024, Schneider2025}, stripped-star and white-dwarf binaries \citep[\eg][]{Li2023, Rubio2025}, and the onset of contact phases and stellar mergers \citep[\eg][]{Pols1994, deMink2007, Claeys2011, Mennekens2017, Henneco2024a}. 

Efforts have been made to obtain more accurate mass-transfer rates in 1D binary evolution calculations \citep[\eg][]{Pavlovskii2015, Marchant2021, Cehula2023, Ivanova2024}. When determining the stability of mass transfer, which dictates whether mass transfer leads to contact phases, most attention has been given to the response of the donor star \citep[\eg][]{Hjellming1989a, Hjellming1989b, Deloye2010, Ge2010, Woods2011, Ge2010b, Pavlovskii2015, Ge2015, Ge2020, Ge2023, Temmink2023, Temmink2025, Ge2024, Ivanova2024}; however, the response of the accretor to mass transfer is similarly important as most recently shown by, \eg, \citet{Ercolino2024} and \citet{Henneco2024a}. 

The efficiency of mass transfer, that is, the fraction of the mass transferred by the donor that is accreted by the companion, has been studied in detail by, \eg, \citet{deMink2007} and \citet{Claeys2011}. In these works, the mass-transfer efficiency was treated as a parameter, and their resulting synthetic populations of binary stars were compared to observations of double-lined spectroscopic binaries and Type IIb supernova progenitors. Using the rapid binary population synthesis code \texttt{COMPAS} \citep{Stevenson2017, Vigna-Gomez2018}, \citet{Willcox2023} varied the mass-transfer efficiency and specific angular momentum carried away by non-accreted matter and showed how this affects the occurrence of stable mass transfer in a population of binary systems with initial primary masses between 5 and $100\,\msun$. Recently, \citet{Lechien2025} analysed a sample of post-mass-transfer binaries with Be and subdwarf OB components and concluded that mass transfer must be relatively efficient to explain the properties of these systems. Yet, the mass-transfer efficiency remains unconstrained across the large range of orbital separations, mass ratios, and evolutionary stages of binary systems. More recently, \citet{Chen2026} demonstrated how the mass-transfer efficiency and stability affect the predicted merging binary black hole mass function.

In \citet{Henneco2024a} (erratum: \citealt{Henneco2024corr}), hereafter \citetalias{Henneco2024a}, we used 1D binary evolution models to explore which binary systems with initial primary masses between 0.8 and $20.0\,\msun$ evolve into contact configurations and which of those likely lead to stellar mergers. A better understanding of the progenitors of these contact phases and stellar mergers is a vital first step in unravelling the physical properties of their products and the merging phases themselves (see \citealt{Schneider2025review} for a recent review). The adopted and commonly used assumptions regarding the spin-up and subsequent quenching of accretion in the binary evolution models from \citetalias{Henneco2024a} led to relatively low overall mass-transfer efficiencies, especially in binaries that initiate mass transfer after the end of the main sequence (MS). Moreover, it was found that with such low mass-transfer efficiencies, the non-accreted matter could not be ejected from almost all binary systems, prompting the question of whether this excess mass could lead to inspiral and mergers, or form a circumbinary disk or shell. In the first part of this work, we explore what happens to the evolutionary outcomes of the models from \citetalias{Henneco2024a} when we impose fully conservative mass transfer and make a detailed comparison between the two grids of models. 

In the second part of this work, we focus on binary systems during or after stable mass transfer. As a result of mass transfer, the initially more massive and, hence, more evolved star can become the less massive component. When binaries are observed at this stage of their evolution, be it during or after mass transfer, they are classified as Algol-type systems \citep{Paczynski1971, Pustylnik1998}, hereafter referred to as Algols. Another result of mass transfer is that the donor star transfers its hydrogen-rich envelope and becomes a stripped star. Depending on their mass, such binary stripped stars can appear as subdwarfs (${\lesssim}2\,\msun$; \citealt{Heber2016}, \citealt{Gotberg2018}) or Wolf-Rayet stars (${\gtrsim}10\,\msun$; \citealt{Crowther2007}), with those with masses in-between often referred to as stripped helium stars or intermediate-mass stripped stars \citep{Drout2023, Yungelson2024, Ludwig2025}. The accretor, spun up to (near-)critical rotation rates, can be observed as Be stars \citep[\eg][]{Raguzova2005, Liu2006, Peters2008, Wang2017, Wang2021, Milone2018, Hastings2020, Hastings2021, WangChen2020, WangChen2023}, although they may end up with rotation rates well below critical after mass transfer ceases \citep{Wang2026}. Since both Algols and stripped-star binaries are direct products of mass transfer, they are ideal objects to probe our current mass-transfer models \citep[see \eg][]{Sen2022, Lechien2025}. Using the grid of binary evolution models from \citetalias{Henneco2024a} and our current grid, both with different assumptions regarding the mass-transfer efficiency, we create synthetic populations of binary stars and compare them to observed Algols and stripped-star binaries to understand which model better matches observations.

This work is organised as follows: we give an overview of our methods to create the grid of detailed 1D binary evolution models and synthetic population of binary stars in Sect.~\ref{sec:methods}. In Sect.~\ref{sec:contact_tracing}, we trace which binary systems enter contact and may merge as a result, and compare to \citetalias{Henneco2024a}. Section~\ref{sec:stable-mt} presents our results on the orbital properties and masses of Algols and stripped-star binaries from our grid and compares them to observations. We discuss our results in Sect.~\ref{sec:discussion} and conclude in Sect.~\ref{sec:conclusions}.

\section{Methods}\label{sec:methods}

Following \citetalias{Henneco2024a}, we computed a grid of 5802 1D binary evolution models with \texttt{MESA} \citep[r12778;][]{Paxton2011, Paxton2013, Paxton2015, Paxton2018, Paxton2019}. The setup of this grid is quasi-identical to that of \citetalias{Henneco2024a}, that is, the sampling of the initial primary masses $M_{1,\,\mathrm{i}}$ and initial mass ratios $q_{\mathrm{i}} = M_{2,\,\mathrm{i}}/M_{1,\,\mathrm{i}}$ are exactly the same. We have slightly lower resolution in the initial binary separation $a_{\mathrm{i}}$ for the initially widest systems than in \citetalias{Henneco2024a}. This was done since most models in this region of the initial binary parameter space do not engage in mass transfer anyway. Save from a few mass-transferring models at the boundary between mass transfer and no mass transfer, this setup allows for one-to-one comparisons between models of the two grids. In the grid from \citetalias{Henneco2024a}, mass transfer could become non-conservative through rotation-limited accretion \citep{Langer2003a}, \ie, accretion stops, and the non-accreted matter is assumed to be lost to infinity when the accretor in our models reaches 97\% of critical surface rotation \citepalias{Henneco2024a}. In some individual models experiencing numerical difficulties, \citetalias{Henneco2024a} limits the accretion timescale $\tau_{\mathrm{acc}}$ to $0.1$ times the accretor's global thermal timescale $\tau_{\mathrm{KH}}$. In our new grid, we enforced fully conservative mass transfer -- except for mass loss through stellar winds -- by not modelling rotation and not limiting the accretion rate to alleviate numerical issues. As a result, we did not model the effect of tides either. Apart from these differences, we used the same physical and numerical assumptions as in \citetalias{Henneco2024a}, which we summarised in the section below. In the remainder of this work, we will refer to the models from \citetalias{Henneco2024a} as the `rotation-limited accretion' (RLA) models, even though some models do not reach this regime thanks to strong tidal interactions, and the models of this work will be referred to as the `fully conservative mass transfer' (FCMT) models.

\subsection{Adopted stellar and binary physics}\label{sec:adopted_physics}
We considered binary systems with initial primary masses $M_{1,\,\mathrm{i}}$ between $0.8$ and $20.0\,\msun$ and initial secondary masses $M_{2,\,\mathrm{i}} \geq 0.5\,\msun$. Initial mass ratios $q_{\mathrm{i}}$ were taken between 0.1 and 0.97. The initial binary separation sampling covers Case-A (donor star on the main sequence), Case-B (post-main-sequence donor star before central helium ignition), and Case-C (post-MS donor star after central helium ignition) mass transfer, up to initial separations $a_{\mathrm{i}}$ that do not result in mass transfer. We further subdivide Case-B mass transfer into early Case-B (Case-Be) and late Case-B (Case-Bl). This division is based on the donor star's envelope; for early Case-B systems, the donor star is on the Hertzsprung gap and has a predominantly radiative envelope, while donors in late Case-B systems have developed predominantly deep convective envelopes \citepalias[see][]{Henneco2024a}.  Both binary components were evolved in our models.

We initialised each binary component from a precomputed zero-age main-sequence model. We computed the models at solar-like metallicity, that is, $Z = 0.0142$ and $Y = 0.2703$ \citep{Asplund2009}, and used a combination of the OPAL \citep{Iglesias1993, Iglesias1996} and \citet{Ferguson2005} opacity tables suitable for the \citet{Asplund2009} chemical composition. We used the \texttt{approx21} nuclear network and disabled \texttt{MESA}'s implicit hydrodynamic solver (\ie all models are hydrostatic at all times).

A combination of prescriptions was used to compute mass loss through stellar winds. This combination consisted of the \citet{Vink2000}, \citet{Vink2017}, and \citet{Sander2020} prescriptions for hot stars and the \citet{Reimers1975}, \citet{Bloecker1995}, and \citet{Nieuwenhuijzen1990} prescriptions for cool stars. We refer to \citetalias{Henneco2024a} for further details.

We used mixing length theory \citep{Bohm-vitense1958, Cox1968} for convective zones with a mixing length parameter $\alpha_{\mathrm{mlt}} = 2.0$ \citep{Paxton2013} and the Ledoux criterion to determine the local stability against convection. Semi-convective mixing and thermohaline mixing were included with efficiencies of $\alpha_{\mathrm{sc}} = 10.0$ \citep{Schootemeijer2019} and $\alpha_{\mathrm{th}} = 1.0$ \citep{Marchant2021}. We included convective boundary mixing (CBM) using the step-overshoot scheme, allowing convective hydrogen burning cores to extend $0.20 H_{P}$ \citep{Martinet2021}, and non-burning convective envelopes $0.05 H_{P}$ \citep{Angelou2020} beyond the convective boundary set by the Ledoux criterion, with $H_{P}$ the pressure scale height. We used the exponential overshoot scheme to include CBM for $0.005 H_{P}$ beyond the convective boundary of convective cores beyond the main sequence \citep{Marchant2021}. Because our current models do not include rotation, there is less overall envelope mixing than in the models of \citetalias{Henneco2024a}. This can affect our models' surface abundances. However, both the current models and those in \citetalias{Henneco2024a} apply additional smoothing to the chemical gradients at the surfaces of accretors to promote convergence during mass-transfer phases.

The onset and rate of Case-A mass transfer were computed via \texttt{MESA}'s \texttt{contact} scheme, which is a combination of the \texttt{roche\_lobe} scheme for semi-detached systems and the scheme from \citet{Marchant2016} for systems where both components (over)fill their respective Roche lobes. The former scheme computes the mass-transfer rate such that the donor star remains inside its Roche lobe. For post-MS donor stars, we used the \citet{Kolb1990} scheme, which considers the optically thick and thin regions of the donor star's envelope. All the above schemes were solved implicitly. An important difference with the treatment of mass transfer compared to that in \citetalias{Henneco2024a} is that in the current models, we allowed the model to continue evolving -- by switching to the \texttt{contact} scheme -- once the accretor star (over)fills its Roche lobe regardless of the evolutionary state of the donor star. Although the \texttt{contact} scheme is primarily meant for Case-A systems, we found that Case-Be still have relatively dense, radiative envelopes, justifying the scheme's usage to an extent. Moreover, since we expected more Case-Be systems to reach contact with fully conservative mass transfer, we were interested in their further evolution. However, the results past the onset of contact in Case-Bl and -C systems should be interpreted with caution.

\subsection{Stopping conditions}
Each model was terminated by default at core carbon (C) exhaustion, that is, when the central helium (He) and C mass fractions fall below $10^{-6}$ and $10^{-2}$, respectively, in either component. The models were terminated before core-C exhaustion when (1) L$_{2}$-overflow occurred while using the \texttt{contact} scheme, (2) the accretor overfilled its Roche lobe by more than its own Roche-lobe radius $R_{\mathrm{RL,\,2}}$ when using the \citep{Kolb1990} scheme, (3) the mass-transfer rate exceeded $10\,\msun \mathrm{yr}^{-1}$, or (4) there is reverse mass transfer from the secondary to the primary star.

\subsection{Population synthesis}\label{subsec:methods-Popsyn}
Each binary model is assigned a birth probability $p_{\mathrm{birth}}$ as in \citetalias{Henneco2024a}. These $p_{\mathrm{birth}}$ used the \citet{Kroupa2001} initial mass function for the initial primary mass $M_{1,\,\mathrm{i}}$, and uniform distributions for the initial mass ratio $q_\mathrm{i}$ and logarithmic initial orbital period $\log P_\mathrm{i}$. Furthermore, we employed a constant star formation rate and convolved this with the time our models spend in certain evolutionary phases (if applicable). 

In our populations, we also consider eclipsing binaries. For a binary system with orbital separation $a$, and stellar radii $R_1$ and $R_2$, the system shows eclipses if the inclination angle $i$ at time $t$ is
\begin{equation}
    i(t) \geq \cos^{-1}\left(\frac{R_1(t) + R_2(t)}{a(t)}\right).
\end{equation}
For randomly oriented binary orbits, the probability of observing a model as an eclipsing binary is thus given by
\begin{equation}\label{eq:prob_with_eclipse}
    p_\mathrm{ec}(t) = \frac{R_1(t) + R_2(t)}{a(t)} p_{\star},
\end{equation}
with the stellar radius being replaced by the Roche radius for semi-detached systems, and all the variables except the observation probability $p_{\star}$ are time-varying. This observation probability is computed through
\begin{equation}
    p_{\star} = p_\mathrm{birth}\int_{t_\mathrm{i}}^{t_\mathrm{f}}C_\mathrm{t}dt\quad ,
\end{equation}
where $C_\mathrm{t}$ is a constant for our assumption of a constant star formation rate. Because the stellar radii and the orbital separation vary with time, this probability is applied to each model timestep. For general information on constructing arbitrary stellar populations, the reader is refereed to \citet{Izzard2018a}.

For our study, we define Algols as systems that have a `theoretical' mass ratio $q_\mathrm{th}$ greater than one, where $q_\mathrm{th}$ is defined as the ratio of masses of the initial secondary over the initial primary star. Since we do not consider reverse mass transfer, this is equivalent to $q_\mathrm{th}=M_\mathrm{accretor} / M_\mathrm{donor}=M_2/M_1.$ In our models, binary stars undergoing mass transfer as well as post-mass-transfer binary systems are classified as Algols whenever $q_\mathrm{th}>1$. We define stripped-star binaries as post-mass-transfer systems in which the primary star has lost its hydrogen envelope, \ie, has a surface helium fraction of $Y_\mathrm{surf}>0.6$, and an effective temperature, $T_\mathrm{eff}$, hotter than the zero-age main sequence (ZAMS) at the current luminosity of the donor star.

\section{Contact binaries, classical common envelopes, and stellar mergers}\label{sec:contact_tracing}

\begin{figure*}
\centering
  \includegraphics[width=18cm]{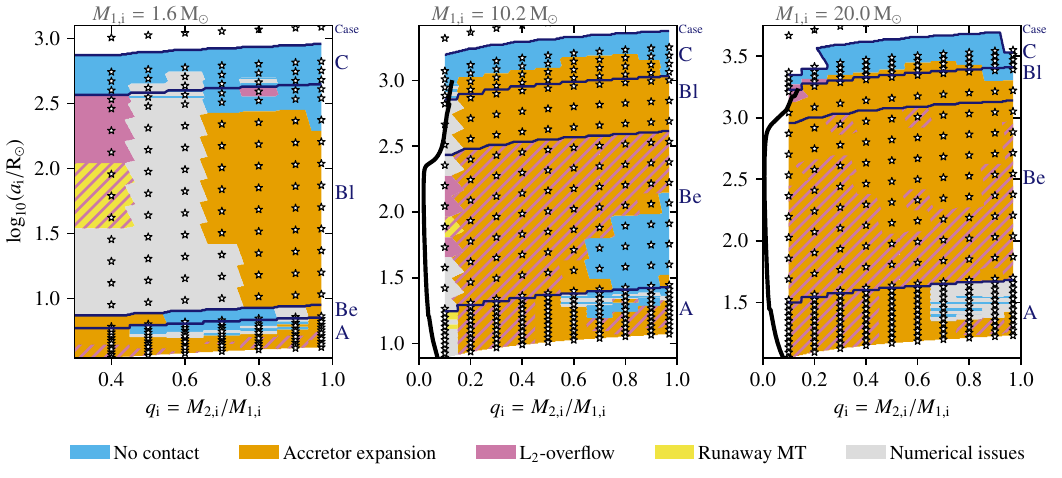}
     \caption{Occurrence of contact phases for models with initial primary masses $M_{1,\,{\mathrm{i}}} = 1.6$ (left), $10.2$ (middle), and $20.0\,\msun$ (right) in the initial mass ratio--separation plane. Each star symbol represents the initial mass ratio and binary separation of a computed \texttt{MESA} model. The initial mass ratio--separation plane is separated into zones with different initial mass-transfer cases. Following \citetalias{Henneco2024a}, each model is labelled with its primary outcome by different solid background colours (`No contact', `Accretor expansion', `Runaway MT', and `$\mathrm{L}_{2}$-overflow') and potentially an ancillary information is added as hatchings. The diagonal hatching indicates `$\mathrm{L}_{2}$-overflow' and the horizontal hatching indicates the `Assumed no contact despite numerical issues' outcome. The solid-black line indicates the lowest initial mass ratio below which binary systems are theoretically Darwin unstable at the onset of mass transfer, following \citet{Rasio1995a}.}
     \label{fig:contact_tracing_three}
\end{figure*}

Figure~\ref{fig:contact_tracing_three} shows the occurrence of contact phases for 3 of the 23 initial primary masses considered in this work ($M_{1,\,\mathrm{i}} = 1.6\,\msun$, $10.2\,\msun$, and $20.0\,\msun$). Similar figures for the other initial primary masses are shown in Appendix~\ref{app:other_contact_tracing}. In contrast to the models in \citetalias{Henneco2024a}, there are no cases of highly non-conservative mass transfer where the excess mass cannot be ejected from the binary and also no occurrences of tidally-driven contact as the models in this work do not rotate. Models in which mass transfer occurs after thermal pulses on the asymptotic giant branch (AGB) are labelled `MT after TP' and their outcome is uncertain, because we do not model thermal pulses accurately with our current setup (see section~2.1.2 in \citetalias{Henneco2024a}). Following \citetalias{Henneco2024a}, we label models that make it to the end of core-He burning and are, therefore, strong candidates for avoiding contact but encountered numerical difficulties before they could reach the end of core-C burning, as `Assumed no contact despite numerical issues'. These models are incorporated in the regular `No contact' category in the incidence computations in Sect.~\ref{sec:incidences}. In this work, we only trace those contact phases that are encountered because of mass transfer from the initially more massive star (primary). This does not mean, however, that these systems cannot enter contact phases later on in their evolution, for example, after the primary has gone supernova or during episodes of reverse mass transfer. 

We consider the following mechanisms leading to contact phases \citepalias[\cf section~3 in][]{Henneco2024a}:

\paragraph{Accretor expansion.} These systems enter a contact phase because of the expansion of the accretor during mass transfer. This may happen if the accretor is brought out of thermal equilibrium and expands on the thermal (Kelvin-Helmholtz) timescale $\tau_{\mathrm{KH}}$. Alternatively, contact may happen because of nuclear-timescale ($\tau_{\mathrm{nuc}}$) expansion of the accretor. The technical definition for models labelled as `Accretor expansion' is that the accretor fills its Roche lobe simultaneously with the already Roche-lobe (over)filling donor star.

\paragraph{Runaway MT.} When the donor star increasingly overfills its Roche lobe, either through rapid expansion, too slow contraction, or orbital decay, mass transfer can become unstable and enter a runaway situation, leading to a contact phase. Since our \texttt{MESA} setup does not allow us to simulate dynamical-timescale mass transfer, we predict the onset of runaway mass transfer phases by a set of conditions: (1) the donor star's radius is larger than its Roche lobe radius, that is, $R_{1} > R_{\mathrm{RL},\,1}$, and the values of $R_{1}$ and $R_{\mathrm{RL},\,1}$ diverge; (2) the mass-transfer rate exceeds the thermal mass-transfer rate $\dot{M}_{\mathrm{KH}} = M_{1}/\tau_{\mathrm{KH}}$; and (3) the second derivative of $\log \dot{M}_{\mathrm{trans}}$ is positive, signalling accelerated growth of the mass-transfer rate $\dot{M}_{\mathrm{trans}}$. We note that condition (1) is not applied to Case-A mass-transfer phases because the \texttt{roche\_lobe} scheme, used for semi-detached binaries with MS donor stars, ensures that $R_{1} \leq R_{\mathrm{RL}}$. 

\paragraph{$\mathrm{L}_{2}$-overflow.} When mass is lost from a binary system through its outer Lagrange point $\mathrm{L}_{2}$ or $\mathrm{L}_{3}$, this is typically accompanied by a significant loss of angular momentum (depending on the density of stellar material at these points and the outflow velocity, see \citealt{Marchant2021}). When $\mathrm{L}_{2}$-overflow occurs in a contact binary, this typically leads to a stellar merger or a common envelope phase. In our models, we track whether either of the binary components of a contact binary becomes larger than the volume-equivalent radius of the $\mathrm{L}_{2}$ lobe, $R_{\mathrm{L}_{2}}$, using the fitting formula from \citet{Marchant2016}. In semi-detached binary systems, we do the same using the appropriate fitting formulae from \citet{Misra2020}. Although \texttt{MESA} assumes the binary is still semi-detached when $R_{1} \sim R_{\mathrm{L}_{2}}$, a donor star of such proportions would in reality have engulfed its companion and caused the onset of a classical common envelope phase.

The most striking difference between the contact phases in Fig.~\ref{fig:contact_tracing_three} and those of \citetalias{Henneco2024a} is the much higher incidence of contact triggered by accretor expansion. In the RLA models of \citetalias{Henneco2024a}, this mechanism mostly operated in Case-A systems, whereas in the present FCMT models it dominates not only Case-A but also much of the Case-B and even Case-C regions of the $(q_{\mathrm{i}}\text{–}\log a_{\mathrm{i}})$ diagram. This results from the efficient mass transfer in FCMT models; in the RLA models, accretion was typically quenched after the accretor had gained only ${\lesssim}3\%$ of its initial mass. As discussed by \citetalias{Henneco2024a}, contact through accretor expansion during Case-A or Case-Be mass transfer occurs while both stars are on the MS or in the Hertzsprung gap (HG). Because both stars are relatively compact, these systems evolve into (`peanut-shaped') contact binaries with components of comparable size. In contrast, when contact is reached during Case-Bl or Case-C mass transfer, the donor is already a (super-)giant. For the accretor to fill its Roche lobe in such wide systems, it must also expand to giant dimensions, producing a contact configuration containing two giant-like stars. These systems differ from classical common-envelopes (CEs) , in which only one component is a giant while the other component is relatively compact (\eg an MS star or compact object). Following \citet{Roepke2023}, we therefore classify them as double-core CEs. Notably, such systems were absent from the binary models with RLA of \citetalias{Henneco2024a}, highlighting another important consequence of FCMT.
In the following sections, we discuss the results for each initial primary mass shown in Fig.\ref{fig:contact_tracing_three}. An equivalent of the outcome table presented in Appendix~G of \citetalias{Henneco2024a} is provided in Appendix~\ref{app:table}.

\subsection{Initial primary mass of $10.2\,\msun$}\label{subsec:Initial10.2M}
Binary systems with the lowest initial mass ratios run into numerical issues for the majority of initial binary separations (middle panel of Fig.~\ref{fig:contact_tracing_three}). The accretors in these systems have convective envelopes, which often complicates numerical convergence. The expansion timescales of these accretors, $\tau_{R/\dot{R}} \equiv R/\dot{R}$, approach the stars' dynamical timescale $\tau_{\mathrm{dyn}}$, and the model enters a regime which goes beyond our \texttt{MESA} setup's capabilities. The intricacies of such rapid expansion phases during mass accretion have been the focus of recent work by \citet{Lau2024}, \citet{Zhao2024}, and \citet{Schuermann2024}. Looking at the results of other $M_{1,\,\mathrm{i}}$ just above and below $10.2\,\msun$ (Appendix~\ref{app:other_contact_tracing}), we find that it is likely that these models would either reach contact through runaway mass transfer or accretor expansion. This is also in line with what is found for the equivalent models in \citetalias{Henneco2024a}. Moreover, binary systems with initial mass ratios lower than the black solid line in Fig.~\ref{fig:contact_tracing_three} are Darwin unstable at the onset of mass transfer and expected to merge \citep[\cf][]{Rasio1995a}.

For Case-A binaries, we find similar behaviour as for the equivalent models from \citetalias{Henneco2024a}. This is no surprise, given that tides efficiently synchronise the accretors in the Case-A models of \citetalias{Henneco2024a}, resulting in effectively conservative mass transfer. We recover the same tendency in our current models to experience $\mathrm{L}_{2}$-overflow when $q_{\mathrm{i}} \leq 0.45$ and $q_{\mathrm{i}} \geq 0.65\text{--}0.75$, which results in stellar mergers. Contrary to the RLA models in \citetalias{Henneco2024a}, the FCMT Case-A systems at $\log(a_{\mathrm{i}}/\rsun) \approx 1.25$ and $0.35 < q_{\mathrm{i}} \lesssim 0.55$ do not avoid contact by accretor expansion. Due to the weaker tidal torques at larger $a_{\mathrm{i}}$, the accretors in the equivalent RLA Case-A models in \citetalias{Henneco2024a} spin up and accretion is quenched, whereas the non-rotating accretors in our current models continue accreting enough to reach contact. 

At $q_{\mathrm{i}} \gtrsim 0.55$, we find several models that avoid contact and form stripped-star binaries. Some of these models initiate Case-C mass transfer when the relatively thin hydrogen layer of the stripped-star primary expands after core-He exhaustion. Due to numerical difficulties during this Case-C mass-transfer phases, not all models make it to core-C exhaustion. However, from the models that do make it that far, and the discussion in \citetalias{Henneco2024a}, we do not expect these mass-transfer phases to lead to contact. In some of these models that avoid contact, reverse mass transfer phases occur after the secondary star expands during its post-MS evolution. Because of the numerical difficulties that arise in modelling such reverse mass-transfer phases with our current setup, we disregard them in this work, as was done in \citetalias{Henneco2024a}. 

It is interesting to explore why binaries in this part of the initial binary parameter space avoid contact despite mass transfer being fully conservative. These systems are initially wider, resulting in a larger Roche lobe to fill for the accretor. Moreover, their initial mass ratios are closer to 1, meaning that the orbit starts to widen sooner after the onset of mass transfer than in systems with lower initial mass ratios. Lastly, we also find that the accretors in these systems, with initial masses between $6.1$ and $9.9\,\msun$, relatively quickly regain thermal equilibrium after the onset of mass transfer. After an initial expansion phase, their radii stay roughly the same during the remainder of the initial Case-A mass transfer phase.

We note that the division line between Case-A and Case-Be mass transfer lies at somewhat lower $\log (a_{\mathrm{i}}/\rsun)$ compared to the one in figure~7 of \citetalias{Henneco2024a}. This is not related to tides or the efficiency of mass transfer, but to the setup of our models. Since our current models do not rotate, they have, in general, smaller radii than their rotating counterparts in the grid with RLA models. This can prevent the primary stars in the current models from filling their Roche lobe, while their rotating counterparts do fill their Roche lobe and initiate mass transfer.

In stark contrast with the RLA models in \citetalias{Henneco2024a}, where the accretors spin up to their critical rotation rates and effectively stop accreting, the majority of the FCMT Case-Be models in this work form contact binaries. \texttt{MESA}'s \texttt{contact} scheme predicts that virtually all of these systems eventually experience $\mathrm{L}_{2}$-overflow, which we assume in Case-Be systems causes them to merge. Even if we were to ignore the predictions of the \texttt{contact} scheme and base ourselves on the state of Case-Be contact binaries at the onset of contact, it would be safe to assume that a merger follows due to the rapid, thermal timescale expansion of both the donor and accretor. 

The Case-Be binaries roughly below the line connecting $(q_{\mathrm{i}};\,\log a_{\mathrm{i}}/\rsun) \approx (0.65;\,1.75)$ and $(q_{\mathrm{i}};\,\log a_{\mathrm{i}}/\rsun) \approx (0.97;\,2.10)$ avoid contact and become stripped-star binaries. We need to explore two trends to explain why these systems avoid contact: one in $q_{\mathrm{i}}$ and one in $a_{\mathrm{i}}$. Accretor stars in systems with lower mass ratios have smaller Roche lobes than those in systems with higher mass ratios and the same separation. Moreover, during conservative mass transfer, the orbit, and hence both components' Roche lobes, shrink as long as $q = M_{2}/M_{1} < 1$. In systems with $q_{\mathrm{i}}$ closer to 1, the accretor initially has a larger Roche lobe and the orbital separation does not decrease as much as in systems with lower initial mass ratios before reaching $q = 1$. All this allows the accretors in higher $q_{\mathrm{i}}$ systems to expand to larger sizes without filling their Roche lobe. The trend in $a_{\mathrm{i}}$ relates to the mass-accretion rate. We find that the mass-accretion rate increases with increasing (initial) orbital separation $a_{\mathrm{i}}$. Higher accretion rates lead to more rapid and more extreme expansion of the accretors, making them more prone to filling their Roche lobe and initiating contact. Donor stars in initially wider systems have larger radii and, hence, shorter thermal timescales. Case-Be mass transfer happens on this thermal timescale, which explains the higher mass-transfer and accretion rates (equivalent for conservative mass transfer). This picture is consistent with what is described in \citet{Schneider2015}. 

We find that all the Case-Bl systems, as well as some of the Case-C systems, with $q_{\mathrm{i}} \geq 0.15$ reach contact through accretor expansion. As briefly described above, these systems will enter a double-core CE phase, which may or may not lead to a merger -- a question which can only be answered via dedicated 3D simulations. Due to the numerical issues encountered in the equivalent models in \citetalias{Henneco2024a}, we cannot, at present, compare them to our current fully conservative models. At even higher $a_{\mathrm{i}}$, we see that Case-C systems avoid contact, which is counter-intuitive at first. \citet{Henneco2024a} and \citet{Ercolino2024} have shown that Case-C mass transfer can be stabilised by inefficient mass transfer, which leads to less orbital shrinkage and earlier (in terms of the evolution of $q$) orbital widening. Looking at these models in detail, we see that the donor stars in these systems never formally fill their Roche lobe and only initiate optically thin mass transfer (see Sect.~\ref{sec:adopted_physics} and \citealt{Kolb1990}), before shrinking again when they ignite carbon in their cores. The resulting mass-transfer rates are extremely low, $\dot{M}_{\mathrm{trans}} \sim 10^{-13}\msun\mathrm{yr}^{-1}$. Their MS companions are virtually unaffected by accretion, avoiding the onset of contact.

Lastly, we note that due to the decreased resolution in $a_{\mathrm{i}}$ for the initially widest systems (see Sect.~\ref{sec:methods}) in the FCMT grid compared to the RLA grid from \citetalias{Henneco2024a}, the line dividing models engaging in mass transfer and those not engaging in mass transfer is slightly shifted upwards (from an average over all $q_{\mathrm{i}}$ of $\log\left(a_{\mathrm{i}}/\rsun\right) = 3.25$ for the RLA models to $\log\left(a_{\mathrm{i}}/\rsun\right) = 3.29$ for the FCMT models). This is is the case for all initial primary masses in this work.

\begin{figure*}
\centering
  \includegraphics[width=16cm]{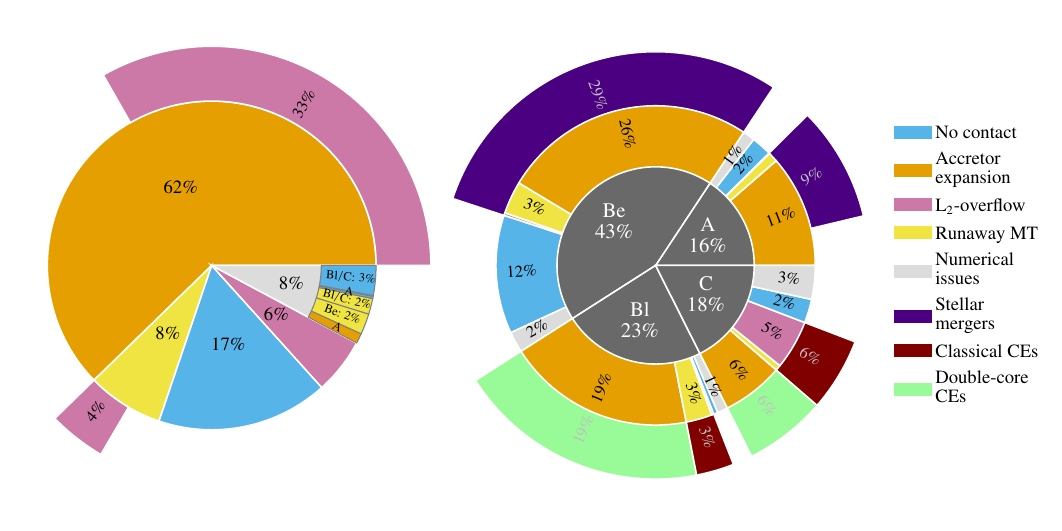}
     \caption{Sunburst charts showing the percentages of contact tracing outcomes for mass-transferring binary systems with $M_{1,\,\mathrm{i}} \in [4.8;\,20.8]\,\msun$. Outcomes with less than $1\%$ are not labelled. The inner level of the sunburst chart on the left shows the principal outcomes of the binary evolution (`Accretor expansion', `Runaway MT', `$\mathrm{L}_{2}$-overflow', `No contact') and the outer level shows which binary models experience $\mathrm{L}_{2}$-overflow as an ancillary outcome. Following the classification scheme in \citetalias[][Table E.1]{Henneco2024a}, models experiencing numerical issue are assigned expected outcomes based on their position in the initial mass ratio-separation plane. The inner level of the sunburst chart on the right shows the percentages of first mass-transfer cases. The middle level indicates the principal outcomes of our models, and the outer level the lower limits of the stellar merger and classical and double-core CE fractions.}
     \label{fig:pie_charts}
\end{figure*}

\subsection{Initial primary mass of $20.0\,\msun$}
We see that the evolutionary outcomes for the models with initial primary masses $M_{1,\,\mathrm{i}} = 20.0\,\msun$ (Fig.~\ref{fig:contact_tracing_three}, right panel) are roughly the same as for the models with $M_{1,\,\mathrm{i}} = 10.2\,\msun$. The first noticeable difference is that we encounter considerably fewer numerical issues. Contrary to the models with $M_{1,\,\mathrm{i}} = 10.2\,\msun$, all considered accretors now have radiative envelopes and are less prone to run into numerical issues when accreting. Looking at the contact tracing results in Fig.~\ref{fig:appendix_contact_tracing1}, we see that these numerical issues indeed disappear between $M_{1,\,\mathrm{i}} = 12.0\,\msun$ and $M_{1,\,\mathrm{i}} = 13.1\,\msun$, which, for smallest initial mass ratios considered here ($q_\mathrm{i}=0.1$), is consistent with the mass at which MS stars have convective envelopes in our setup ($M_\mathrm{i} \lesssim 1.3\,\msun$). The next difference is the lack of Case-Be systems that avoid contact. We find that the more massive accretors in these systems expand more rapidly than lower-mass ones, and that they have less space in their Roche lobe. This last fact stems from the following consideration: at ZAMS, the radius of a star (in this case the accretor) scales as $R_{2}\propto M_{2}^{1/2}$ \citep{Kippenhahn2012}. Its Roche lobe radius $R_{\mathrm{RL,\,2}} \propto a_{\mathrm{orb}}$ and $a_{\mathrm{orb}} \propto (M_{1} + M_{2})^{1/3}$. If we assume a fixed $q$ for our comparison, then $R_{\mathrm{RL}} \propto (qM_{2} + M_{2})^{1/3} \propto M_{2}^{1/3}$. In other words, the radius scales more strongly with mass than the Roche lobe radius. When comparing a more and a less massive binary with the same $q$ at ZAMS, the accretor in the more massive binary is closer to filling its Roche lobe than the accretor in the less massive binary. By looking at the other initial primary mass results in Appendix~\ref{app:other_contact_tracing}, we see that all systems with $M_{1,\,\mathrm{i}} \geq 14.2\,\msun$ do not have Case-Be systems that avoid contact. Next, we see that the line between mass-transferring and non-mass-transferring systems is less well defined for $M_{1,\,\mathrm{i}}  = 20.0\,\msun$ and $M_{1,\,\mathrm{i}} = 8.6\,\msun$ as for other $M_{1,\,\mathrm{i}}$. This is a result of the aforementioned extremely low mass-transfer rates in the widest Case-C systems, which are close to the cut-off mass-transfer rate we defined for our grid, which is $\dot{M}_{\mathrm{trans}} = 10^{-16}\,\msun\mathrm{yr}^{-1}$.

Similarly, as for the systems with $M_{1,\,\mathrm{i}} = 10.2\,\msun$, we find that the contact tracing outcomes for our fully conservative models are in stark contrast with the picture from RLA in \citetalias{Henneco2024a}. Whereas in the former all Case-Be and virtually all Case-Bl systems reach contact, the spin up of the accretors in the latter results in most of these systems avoiding contact, except for those with $q_{\mathrm{i}} \leq 0.15\text{--}0.35$. The mechanism leading to contact is also different in our current models: contact is reached through the expansion of the accretor, whereas runaway mass transfer and $\mathrm{L}_{2}$-overflow are responsible for contact in the models with RLA.

\subsection{Initial primary mass of $1.6\,\msun$}

For the systems with $M_{1,\,\mathrm{i}} = 1.6\,\msun$ (Fig.~\ref{fig:contact_tracing_three}, left panel), we find that, as for the previously described $M_{1,\,\mathrm{i}}$, that the lack of rotation-limited-accretion favours contact by accretion expansion even for initially wide Case-Bl binaries. However, due to the same numerical issues as those described for the equivalent models in \citetalias{Henneco2024a}, we cannot get a full picture of the Case-Bl models. This confirms that these specific numerical issues are not related to the spin-up of the accretors in these models. Lastly, we note the lack of Case-A systems driven into contact by tides. This is a logical consequence of rotation not being modelled in our current models, and is an obvious limitation of our current computational setup. Finally, we see that the fully conservative models have fewer numerical issues during Case-C mass transfer, whereas we could get little information about these systems in the rotation-limited-accretion models from \citetalias{Henneco2024a}. We find that, similarly to what was found for initially wider Case-Bl systems with rotation-limited-accretion, strong stellar winds before the onset of mass transfer, and the fact that, in some cases, mass transfer occurs during the TP-AGB phase, have a stabilising effect on mass transfer. As a result, these systems avoid contact.

\subsection{Incidences of evolutionary outcomes}\label{sec:incidences}

Following \citetalias{Henneco2024a}, we use the birth probabilities of binary stars (Eqs.~3--5 in \citetalias{Henneco2024a}) to compute the incidences for various evolutionary outcomes of mass-transferring binary systems with an initial primary masses of $M_{1,\,\mathrm{i}} \in [4.8;\,20.8]\,\msun$ (Fig.~\ref{fig:pie_charts})\footnote{To compute the initial primary mass dependent part of the birth probability, we integrate the initial mass function between lower and upper boundaries $M_{\mathrm{l}}$ and $M_{\mathrm{u}}$, respectively (see Eq. 3 and 5 in \citetalias{Henneco2024a}). These boundaries are taken everywhere as the midpoints between the initial primary masses. For the $20.0\,\msun$ initial primary mass, we take the half distance to the lower adjacent initial primary mass of $18.4\,\msun$ to set the upper boundary $M_{\mathrm{u}} = 20.8\,\msun$.}. The sunburst chart on the left of Fig.~\ref{fig:pie_charts} shows the incidences of the principal (`Accretor expansion', `Runaway MT', `$\mathrm{L}_{2}$-overflow', `No contact') and ancillary (`$\mathrm{L}_{2}$-overflow') outcomes of binary evolution in our populations of binary stars. As expected from the description in the previous section, we see that the incidence of systems reaching contact, either in the form of a contact binary or double-core CE, because of the expansion of the accretor, increases significantly when mass transfer is fully conservative. This incidence ($62\%$) is almost 6 times as large as for the RLA models in \citetalias{Henneco2024a} ($11\%$). As a result, the incidence of systems avoiding contact is considerably lower with fully conservative mass transfer: $46\%$ with RLA and $17\%$ with FCMT. The incidence of mass-transferring binaries experiencing runaway mass transfer ($8\%$) has roughly halved ($17\%$ with RLA), because the accretors tend to fill their Roche lobe before the onset of unstable mass transfer. The number of systems initiating a CE because of the strong expansion of the donor star, which we track using the onset of $\mathrm{L}_{2}$-overflow in semi-detached binaries (\ie systems that have $\mathrm{L}_{2}$-overflow as their principal outcome) is roughly the same as in the models with RLA: $6\%$ with FCMT and $5\%$ with RLA. Because of the generally easier convergence when not modelling rotation and the rapid onset of contact in many systems, we also find that the incidence of models with numerical issues ($8\%$) has more than halved ($21\%$ with RLA).

Owing to the considerably larger contribution of contact binaries from Case-Be mass transfer, we now see from the sunburst chart on the right side of Fig.~\ref{fig:pie_charts} that the incidence of stellar mergers among mass-transferring binaries with $M_{1,\,\mathrm{i}} \in [4.8;\,20.8]\,\msun$ has increased to ${\geq}38\%$, where it was ${\geq}16\%$ with RLA. Even though we may still underestimate the incidence of stellar mergers because (1) we cannot know for certain how the models with numerical issues would evolve further and (2) we only consider stellar mergers from the first contact phase in case there are multiple (see \citetalias{Henneco2024a}), we argue that this incidence approaches the upper limit of the stellar merger incidence. We infer this from the fact that FCMT clearly seems to favour the onset of contact. Looking only at the Case-A systems, we find that the incidence of stellar mergers remains roughly the same (from $8\%$ with RLA to $9\%$ with FMT), because strong tides prevent these systems from reaching the RLA regime in the models from \citetalias{Henneco2024a} before contact is reached. For the initially wider Case-Bl and Case-C binaries, we find that the incidence of double-core CE systems, $25\%$ in total, dominates over the incidence of classical CEs, which decreased from $19\%$ with RLA to $9\%$ with FCMT. As mentioned before, double-core CEs are completely absent in the models with RLA.

\subsection{Comparison with observed contact binaries}

\begin{figure}
    \centering
    \resizebox{0.85\hsize}{!}{\includegraphics{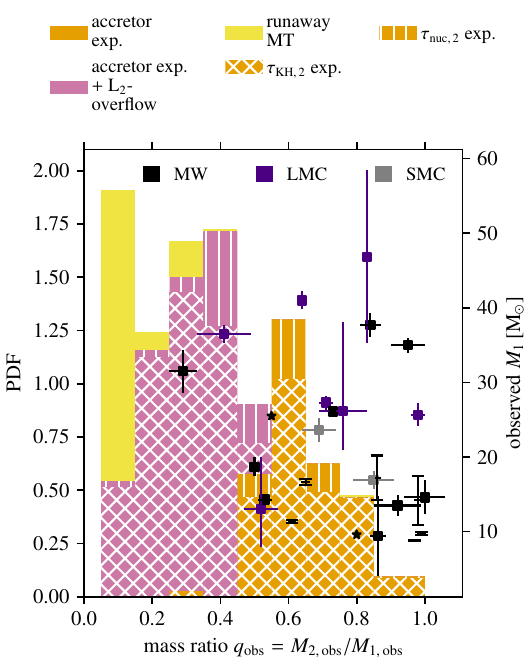}}
    \caption{Probability density function (PDF) of Case-A contact binary systems with initial primary masses $M_{1,\,\mathrm{i}} = 4.8\text{--}20.8\,\msun$ as a function of the observed mass ratio $q_{\mathrm{obs}}$. The filled squares show the observed primary mass and mass ratio of (near-)contact systems and their uncertainties from \citet{Ostrov2001}, \citet{Harries2003}, \citet{Hilditch2005}, \citet{Mahy2020}, and \citet{Janssens2021}, compiled in \citet{Menon2021}. Star symbols indicate observed (near-)contact systems with no reported uncertainties on $q_{\mathrm{obs}}$ and $M_{1}$. Systems with no reported uncertainty on $q_{\mathrm{obs}}$ are indicated with a dash symbol.}
    \label{fig:contact_binaries_histogram}
\end{figure}

In Fig.~\ref{fig:contact_binaries_histogram}, we show the properties of Case-A contact binaries at and the manner in which they reached contact (thermal- or nuclear-timescale expansion, or runway mass transfer) and compare their mass ratios to those of observed massive (near-)contact binaries from \citet{Ostrov2001}, \citet{Harries2003}, \citet{Hilditch2005}, \citet{Mahy2020}, and \citet{Janssens2021}, compiled in \citet{Menon2021}. This is equivalent to the comparison made using the RLA models in Sect.~4.5 of \citetalias{Henneco2024a}. We notice several differences between our current histogram and the one in \citetalias{Henneco2024a}. With FCMT, there is a relatively sharp divide in mass ratio $q_{\mathrm{obs}} = M_{2,\,\mathrm{obs}}/M_{1,\,\mathrm{obs}}$ around $q_{\mathrm{obs}} \approx 0.5$. More equal mass contact binaries seem to avoid $\mathrm{L}_{2}$-overflow, while those with $q_{\mathrm{obs}} \lesssim 0.5$ likely merge via $\mathrm{L}_{2}$-overflow. In the RLA models of \citetalias{Henneco2024a}, this divide is less clear. Nevertheless, this reinforces the conclusion that more unequal mass systems are more likely to merge, thus offering one piece of the explanation of why massive contact binaries are almost exclusively observed with $q_{\mathrm{obs}} \gtrsim 0.5$. We find another major difference in the expansion timescales of the accretors. In the RLA models, systems with $q_{\mathrm{obs}} \gtrsim 0.5$ reach contact predominantly because of the nuclear-timescale expansion of the accretor, often leading to long-lived, potentially observable contact binaries. With fully conservative mass transfer, the accretors are less likely to reach thermal equilibrium, and the onset of contact is driven by their thermal-timescale expansion. As discussed in \citetalias{Henneco2024a}, such systems are more likely to merge relatively quickly or evolve to a semi-detached state once the accretor reaches thermal equilibrium again, making them less likely to be observed as contact binaries. We do still find Case-A contact binaries that form on nuclear timescales with $q_{\mathrm{obs}} \gtrsim 0.5$, but their contribution is lower than in the RLA models. In other words, FCMT, implemented the way it is in our current models, offers a less clear explanation for the observed massive contact binaries with $q_{\mathrm{obs}} \gtrsim 0.5$. One may thus wonder whether FCMT is indeed a suitable model for Case-A mass transfer.

We note that both the mass ratio distribution derived from the FCMT models as the one derived from the RLA models (\citetalias{Henneco2024a}) have a dearth of contact binaries at nearly-equal masses. On the contrary, other works using comparable methods, such as \citet{Menon2021}, predict a considerable probability of observing contact binaries with mass ratios close to 1. Although this might seem discrepant, we reiterate that in Fig.~\ref{fig:contact_binaries_histogram} we show the properties of contact binaries at the moment they reach contact. \citet{Menon2021} consider the evolution through the contact phase and its associated duration when computing their observational probabilities. The scheme they use for contact binary evolution, which is the same one used in this work (see Sect.~\ref{sec:adopted_physics}), tends to lead to systems evolving towards equal-mass configurations. Hence, if we were to create a similar observational probability distribution as in \citet{Menon2021}, we would also expect it to peak at mass ratios close to 1.

\section{Stable Mass Transfer}\label{sec:stable-mt}
We now focus on the evolution of the binary models in the FCMT and RLA grids during and after stable mass transfer. This includes models which eventually reach contact configurations, but do so through stable mass transfer (\eg the `Accretor expansion' models). Such stable mass transfer systems can be observed as Algols and/or stripped-star binaries. In Sect.~\ref{subsec:Mass Transfer Products}, we look at the occurrence of these types of binary systems in the initial binary parameter space and then consider their mass-ratio distributions in Sect.~\ref{subsec:Pop_prop}. In Sect.~\ref{sec:compare_obs_algol_stripped}, we examine the probability of observing these systems in terms of their mass ratio and orbital period. In all of these sections, we compare the FCMT models with RLA models to characterise the impact of the extreme cases of mass-transfer efficiency on the properties on synthetic populations of Algols and stripped-star binaries.

\subsection{Occurrence in the initial parameter space}\label{subsec:Mass Transfer Products}

\begin{figure*}
    \centering
    \includegraphics{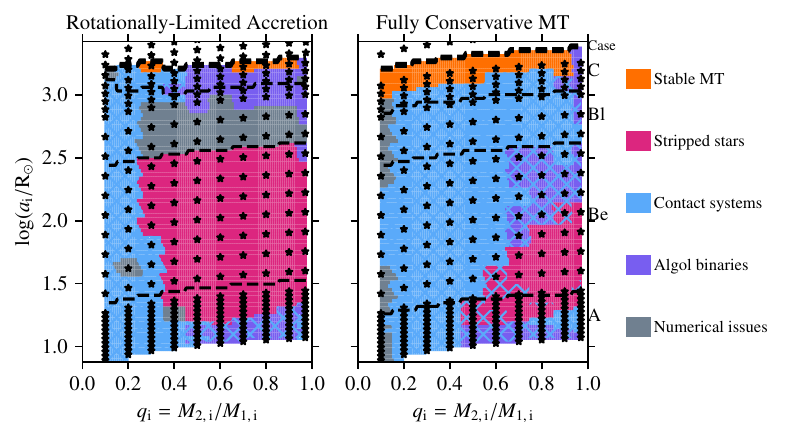}
    \caption{Occurrence of Algols, stripped star binaries, and contact systems with $M_\mathrm{1,\,\mathrm{i}} = 10.2\,\mathrm{M_\odot}$ on the initial mass ratio--orbital separation plane. The left panel shows the RLA models from \citetalias{Henneco2024a}. The right panel shows the FCMT models from our current work. The initial mass ratio--separation plane is separated into zones with different initial mass-transfer cases. The contact systems in this figure represent contact binaries, classical and double-core CEs. Models classified as Algol or stripped-star that also evolve through a contact phase are marked with a hatching.} 
    \label{fig:MT_product_slices}
\end{figure*}
We start by showing the occurrence of mass-transfer products in the initial mass ratio--orbital separation plane in Fig.~\ref{fig:MT_product_slices} for systems with $M_\mathrm{1,\,i}=10.2\,\msun$. Although we focus on the results for $M_\mathrm{1,\,i}=10.2\,\msun$ here, the overall descriptions hold for all systems with $M_{\mathrm{1,\,i}} > 5\,\msun$. The results described here can therefore be generalised for the other $M_{\mathrm{1,\,i}}$ using the figures in Sect.~\ref{sec:contact_tracing} and Appendix~\ref{app:other_contact_tracing} as a guide. The coloured regions in Fig.~\ref{fig:MT_product_slices} indicate the different phases through which a system evolves based on the evolutionary outcome of the nearest \texttt{MESA} model, similar to Fig.~\ref{fig:contact_tracing_three}. As mentioned before, some models evolve into contact while also being observable as Algols or stripped-star binaries during certain phases of their evolution. Such models get a double designation, indicated by the hatching in Fig.~\ref{fig:MT_product_slices}. We find that all models that we classify as stripped-star binaries have their mass ratio reversed, that is $q_{\mathrm{th}}>1$. Even though this automatically puts them in the class of Algols, we simply refer to these systems as stripped-star binaries. The models labelled as Algols are therefore systems in which $q_{\mathrm{th}}>1$, but which do not contain a stripped-star component.

\subsubsection{Case-A Binaries}
For Case-A binaries, we find similar outcomes of mass transfer from both binary model grids (Fig.\ref{fig:MT_product_slices}). Especially the location of short-period Algols in the initial parameter space is notably similar in the two grids. These systems reverse their mass ratio during an initial phase of thermal-timescale mass transfer (\ie before the onset of nuclear-timescale mass transfer, while the donor star has not yet regained thermal equilibrium), as shown by \citet{Sen2022}. Eventually, these systems evolve into contact binaries because of the expansion of the accretor, since accretion is sustained thanks to efficient tidal synchronisation, as discussed in Sect.~\ref{subsec:Initial10.2M}. However, at slightly higher orbital separations ($\log a_\mathrm{i} \gtrsim 1.2\,\mathrm{\rsun}$), we find systems in both grids that undergo Case-AB mass transfer (Case-B mass transfer after a Case-A mass transfer phase) and form stripped-star binaries, some of which go through a contact phase first in the FCMT models. This arises from the higher mass-transfer efficiency in the FCMT models than in the models with RLA, as tidal synchronisation becomes inefficient at larger separations in the latter models.

\subsubsection{Case-Be Binaries}
A striking difference between the two grids is the outcome of models undergoing Case-Be mass transfer. This type of mass transfer is driven by the donor star being out of thermal equilibrium and happens on roughly the donor's thermal timescale. If the accretor is unable to fully relax its thermal structure to its new mass, the ensuing rapid (thermal or faster-than-thermal) expansion of the accretor leads to the formation of contact binaries before the envelope of the donor star can be stripped. Therefore, the FCMT models produce a lower number of stripped-star binaries from Case-Be mass transfer. In contrast, Case-Be models with RLA exclusively develop into stripped-star binaries, except for $q_\mathrm{i}<0.3$. This is because for RLA, accretion is quenched shortly after the onset of mass transfer, as discussed in Sect.~\ref{sec:contact_tracing}, and envelope stripping can continue unhindered by contact phases until the donor star's envelope is completely stripped. However, a majority of these Case-Be models with RLA were classified as `Non-conservative + cannot eject' in \citetalias{Henneco2024a}, which implies that the combined stellar luminosity of the binary components was insufficient to push the non-accreted material to infinity. If such material can induce further angular momentum loss onto the accretor, the binary orbit experiences further shrinkage. Since this loss mechanism is not modelled in \citetalias{Henneco2024a} (or in other similar binary model grid calculations), further evolution and lifetimes of such stripped-stars remain uncertain.

The FCMT models with $a_\mathrm{i}\gtrsim 100\,\rsun$ do not evolve into stripped-star binaries for any given $q_\mathrm{i}$. In this region of the initial parameter space, the accretor stars of models for $q_\mathrm{i}<0.65$ overflow their $\mathrm{L_2}$-point early on during the mass transfer phase and subsequently merge. At initial mass ratios $q_\mathrm{i}>0.65$ and  $a_\mathrm{i}\gtrsim 100\,\rsun$, Case-Be models also eventually reach contact configurations, as indicated by the hatching in Fig.~\ref{fig:MT_product_slices}. However, as discussed in Sect.~\ref{subsec:Initial10.2M}, the accretors in systems with larger initial mass ratios initially have larger Roche lobes, which delays the onset of contact through their expansion. As a result, these systems spend a non-negligible time as Algols before forming contact binaries. 

\subsubsection{Case-Bl and -C Binaries}
In the Case-Bl and Case-C mass-transfer region of the initial parameter space, there is an absence of stripped-star binaries in both grids. We should, however, keep in mind that we cannot be certain about what happens in the Case-Bl region of the models with RLA due to numerical issues, as reported in \citetalias{Henneco2024a}. Nevertheless, we can explain the absence of stripped-star binaries in this region of the initial parameter space as follows: To strip a donor star of its envelope via mass transfer, a sufficiently long mass-transfer phase and a sufficiently high mass transfer rate are required. As discussed in Sect.~\ref{sec:contact_tracing}, the initially widest FCMT Case-C systems have mass transfer with negligibly low mass-transfer rates, which do not lead to any meaningful envelope stripping or mass ratio reversal. Therefore, we label them as `Stable MT' (stable mass transfer). The same holds for some of the initially widest models with RLA. Other models in the Case-Bl and Case-C region tend to reach contact phases, either by runaway mass transfer (models with RLA) or accretor expansion (FCMT models) before they donor star's envelope is fully stripped. The models in both grids that avoid contact transfer enough mass for their mass ratio to become larger than 1, but not enough for the donor star to become stripped. Both Case-Bl and Case-C mass transfer is driven by relatively brief periods of thermal disequilibrium, which are not long enough to fully strip the donor stars.

It is worth restating that even though these systems do not evolve into contact phases now, they can still do so after one of the components, usually the initial primary star, becomes a compact object. As described in detail in \citetalias{Henneco2024a} and Sect.~\ref{sec:methods}, this part of binary evolution is not captured in our models. We also currently do not account for stripped-star binaries resulting from classical or double-core CE phases.

\begin{figure*}
    \centering
    \includegraphics[width=15cm]{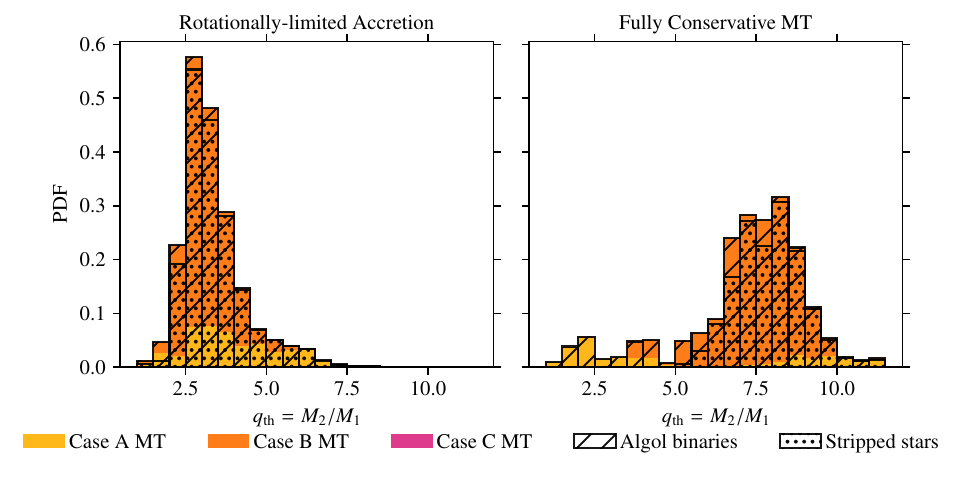}
    \caption{Comparison of 1D probability density functions (PDFs) of the theoretical mass ratio ($q_\mathrm{th}$) of mass-transfer products (Algols and stripped-star binaries) from models with initial primary masses of $M_{1,\,\mathrm{i}} = 4.8\text{--}20.8 \msun$. Both PDFs are normalised to unity. Only models with $q_\mathrm{th}>1$ are shown, \ie they are all Algol binaries. The colours indicate contributions from the various cases of mass transfer, while hatchings represent the type of mass-transfer product.} 
    \label{fig:1D-post-mass-transfer-q}
\end{figure*}

\subsection{Population properties}\label{subsec:Pop_prop}
We now demonstrate the occurrences of stripped-star binaries and Algols on a population level for the RLA and FCMT grids. To this end, we create synthetic binary populations from the two grids following the procedure described in Sect.~\ref{subsec:methods-Popsyn}. Figure~\ref{fig:1D-post-mass-transfer-q} shows the discrete probability density function (PDF) of $q_\mathrm{th}$ for Algols and stripped-star binaries from the two grids, differentiating between Case-A, Case-B, and Case-C mass transfer.

By comparing the two panels of Fig.~\ref{fig:1D-post-mass-transfer-q} (representing the results for the two grids) we find that there is an overall shift of the mean mass-ratio of Algols and stripped-star binaries to higher $q_\mathrm{th}$ for FCMT. In the population based on models with RLA, the $q_\mathrm{th}$-distribution of stripped-star binaries peaks at $q_\mathrm{th} \approx 3$, while in the FCMT population the same distribution has a peak at $q_\mathrm{th} \approx 8$. We further notice that for FCMT, the distributions of stripped-star binaries and Algols are different, with the distribution of Algols extending to much lower mass ratios than that of stripped stars. Most of these low-$q_\mathrm{th}$ Algols underwent stable Case-A mass transfer, and occur at $q_\mathrm{th} \approx 2.0$. The similarity in the $q_{\mathrm{th}}-$distributions of Algols between the two grids can be understood from the fact that the majority of them form through Case-A mass transfer, which generally has higher mass-transfer efficiencies in the RLA models. 

Next, we find that the evolution of the mass ratio in the formation channel of stripped-star binaries is very sensitive to the mass-transfer efficiency. During the formation of such stripped stars, as the entire hydrogen envelope is transferred, the mass of the donor decreases quickly to a value below that of the accretor. These systems reach mass ratios of $q_\mathrm{th}\approx 2-4$. Furthermore, if this entire mass is accreted by the accretor star, as is the case during FCMT, the mass ratio can keep on increasing and might reach extreme values of even $q_\mathrm{th}>10$ in some cases. However, in the RLA grid, most of the accreted matter is lost, which prevents the growth of the accretor, and the mass ratio rarely rises to the extreme values seen in the FCMT grid.

Binaries that undergo Case-C mass transfer contribute negligibly to the post-mass-transfer population. This arises primarily from the short evolutionary timescales characteristic of Case-C mass transfer. Furthermore, some of these models never reach configurations with $q_\mathrm{th}>1.0$ because the Case-C mass-transfer rates are extremely low (see Sect.~\ref{subsec:Initial10.2M}). Since we select for systems with $q_{\mathrm{th}} > 1$ (see Sections~\ref{subsec:methods-Popsyn} and \ref{subsec:Mass Transfer Products}), these systems are excluded.

\begin{figure*}
    \centering
    \includegraphics[width=18cm]{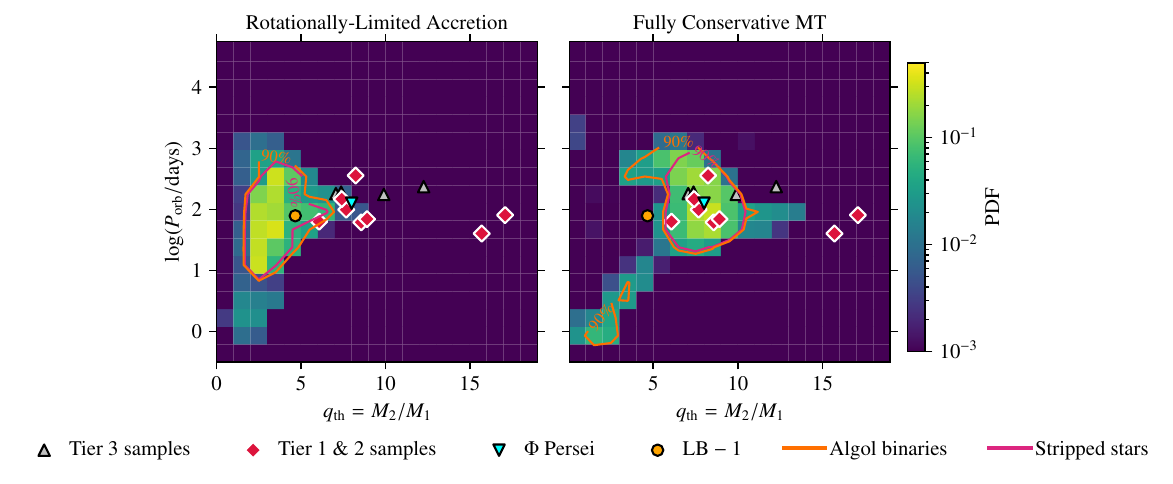}
    \caption{Comparison of 2D probability density functions in the $q_\mathrm{th}$ vs $\log (P_\mathrm{orb})$ plane of post-mass transfer systems from models with $M_{1,\,\mathrm{i}} = 4.8\text{--}20.8\, \msun$. The coloured contours indicate $90\%$ probability contours of Algols and stripped-star binaries. The observations are those of sdOB+Be binaries referenced in Sect.~\ref{sec:compare_obs_algol_stripped}. Definitions of the sample tiers are adapted from \citet{Lechien2025} and represent the degree of confidence in the derived orbital solutions. The properties of $\phi$~Persei, as reported in \citet{Mourard2015}, are indicated separately since it is a prototypical sdOB+Be binary system.}
    \label{fig:2D-post-mass-transfer}
\end{figure*}

\begin{figure*}
    \centering
    \includegraphics[width=15cm]{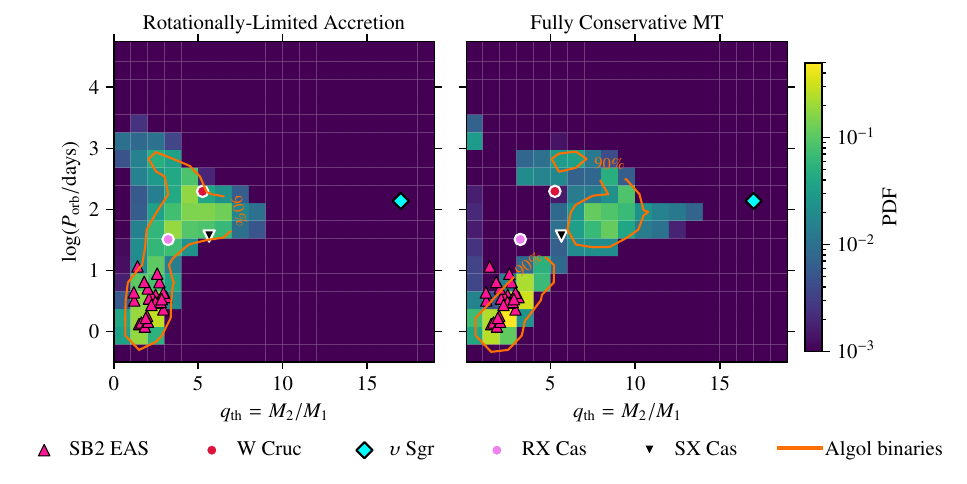}
    \caption{Same as Fig.~\ref{fig:2D-post-mass-transfer}, but for models of mass-transferring binary systems and observations of Galactic short-period semi-detached double-lined eclipsing binaries referenced in Sect.~\ref{sec:compare_obs_algol_stripped}. An additional bias of finding eclipsing binaries as described in Sec.~\ref{subsec:methods-Popsyn} is included while computing the population probability densities.}
    \label{fig:2D-MT-Eclipses}
\end{figure*}

\subsection{Comparison with observed Algols and stripped-star binaries}\label{sec:compare_obs_algol_stripped}
We now compare our synthetic populations with observed Algols and stripped-star binaries. Specifically, we use the observed mass ratios and orbital periods of subdwarf OB-type stars with Be star companions (sdOB+Be binaries) and Galactic Algols. The orbital parameters for the sdOB+Be binaries are from \citet{Mourard2015}, \citet{Klement2022a, Klement2024, Klement2025}, and \citet{Wang2023}, as compiled in \citet{Lechien2025}, and from \citet{Shenar2020b} for the LB-1 system. The orbital parameters for the Algols originate from  \citet{Evans1974}, \citet{Bell1987}, \citet{Bagnuolo1994}, \citet{Lorenz1994}, \citet{Stickland1998}, \citet{Surkova2004}, \citet{Budding2004, Budding2015}, \citet{Terrell2005}, \citet{Hilditch2007}, \citet{Djuraševic2009}, \citet{Surina2009}, \citet{Mahy2011}, \citet{Ibanoglu2013}, \citet{Tuysuz2014}, and \citet{Martins2017}, as compiled in \citet{Sen2022}, and from \citet{Pavlovski2006} for the W Crucis system, \citet{Koubsky2006} and \citet{Gilkis2023} for the $\upsilon$~Sagitarii system, \citet{Andersen1989} and \citet{Mennickent2022} for the RX Cassiopeiae system, and \citet{Struve1944} and \citet{Djurasevic1993} for the SX Cassiopeiae system. 

We compare the orbital parameters of these systems with two-dimensional PDFs for the observation probability $p_{\star}$ in terms of $q_\mathrm{th}$ and $\log (P_\mathrm{orb} /\mathrm{d})$ of binary systems derived from the RLA and FCMT grids in Figs.~\ref{fig:2D-post-mass-transfer} and \ref{fig:2D-MT-Eclipses}. Figure~\ref{fig:2D-post-mass-transfer} shows the 2D PDF for post-mass transfer systems, which we compare with observed sdOB+Be binaries, while Fig.~\ref{fig:2D-MT-Eclipses} shows the 2D PDF for currently mass-transferring binaries, which we compare to observations of Galactic Algol binaries. The majority of these Algols are double-lined eclipsing binaries, hence we need to account for eclipses in our PDFs as discussed in Sect.~\ref{subsec:methods-Popsyn}. Also, the majority of the Algols shown in Fig.~\ref{fig:2D-MT-Eclipses} are semi-detached, which is why we compare them to synthetic populations of currently mass-transferring binaries.

We only consider stripped-star binaries forming via stable mass transfer. A second formation channel for these systems is via successful common envelope evolution \citep{Ivanova2013, Hovis2025, Aguilera2023}. However, since we do not model common envelope evolution in any of the grids, we lack predictions of stripped-star binaries with orbital periods $\lesssim 1$ day. To compensate for this, we do not include observed stripped-star binaries with orbital periods less than a day in our comparison.

In Figs.~\ref{fig:2D-post-mass-transfer} and \ref{fig:2D-MT-Eclipses}, we observe a region of slightly enhanced probability density at $q_\mathrm{th} < 1.0$ and $P_\mathrm{orb}\gtrsim$ 10$^3$\,d. These are core helium-burning red supergiants that have initiated optically thin Case-C mass transfer at rates ${\sim}10^{-13}\msun \mathrm{yr}^{-1}$, as discussed in Sect.~\ref{subsec:Initial10.2M} and Sect.~\ref{subsec:Mass Transfer Products}. 

The PDFs in Fig.~\ref{fig:2D-post-mass-transfer} show clear over-densities for the RLA- and FCMT-based populations at similar orbital periods of around $100\,$\,d, but at different $q_{\mathrm{th}}$. In other words, the stripped-star binaries and post-mass-transfer Algols in both model grids have similar orbital periods, but considerably different mass ratios, clearly showing the effect of the mass-transfer efficiency. We note that for the RLA population, the PDF over-density region is spread out between $q_\mathrm{th}\sim2\text{--}5$, while in the FCMT population, this region is found at $q_\mathrm{th}>6$. Closer inspection shows that the difference in $q_{\mathrm{th}}$-range of the PDF over-densities is mainly caused by differences in the stripped-star binary populations. This can be seen by the contours in Fig.~\ref{fig:2D-post-mass-transfer}. This difference in $q_{\mathrm{th}}$ could already be appreciated from Fig.~\ref{fig:1D-post-mass-transfer-q} as well, and is therefore not unexpected. 

Consistent with the results of \citet{Lechien2025}, we find that the observations support highly conservative mass transfer for the formation of stripped-star binaries: most of the observations are confined within the contour for stripped-star binaries from the FCMT population, and not for those from the RLA population. However, the system LB-1 \citep{Liu2019, Shenar2020b} is an exception, seemingly requiring a lower mass-transfer efficiency to explain its observed properties. This shows that although overall more conservative mass transfer seems to be favoured, there is no reason to expect that a single mass-transfer efficiency value can explain all systems. On the contrary, this efficiency could be system-dependent. 

Two of the observed systems in Fig.~\ref{fig:2D-post-mass-transfer} with $q_{\mathrm{th}} > 15$, HR~6819 \citet{Klement2025} and HR~2142 \citet{Klement2024}, cannot be explained by either extreme of the mass transfer efficiency. Several attempts have been made already to unravel their enigmatic nature \citep[see, \eg][]{Bodensteiner2020, Frost2022, Picco2025}. 

The PDF over densities in both panels of Fig.~\ref{fig:2D-post-mass-transfer} at $P_{\mathrm{orb}} \lesssim 10\,$\,d are caused by the Case-A Algol binaries discussed in Sects.~\ref{subsec:Mass Transfer Products} and~\ref{subsec:Pop_prop}. As mentioned above, Algols are almost exclusively found in semi-detached (\ie mass-transferring) binaries, and therefore ought to be compared to models of mass-transferring binaries (done next).

Fig.~\ref{fig:2D-MT-Eclipses} shows the 2D PDFs of eclipsing post-mass-transfer binary models. The PDFs for both synthetic populations are lower than in Fig.~\ref{fig:2D-post-mass-transfer} due to the combined effect of the observation probability for eclipsing binaries being inversely proportional to the orbital separation (see Eq.~\ref{eq:prob_with_eclipse}) and post-MS mass-transfer phases in initially wider orbits being shorter-lived than for Case-A mass transfer. That said, we find that the effect of the former is relatively minor, since systems with a Roche-lobe filling component show eclipses at a wide range of inclination angles. Nevertheless, the PDFs in Fig.~\ref{fig:2D-MT-Eclipses} are more strongly peaked at lower orbital periods than in Fig.~\ref{fig:2D-post-mass-transfer}.

Comparing our synthetic populations with the aforementioned observational data from Algols compiled in \citet{Sen2022}, we find that at $P_\mathrm{orb}<3\,\mathrm{d}$, the observational data seem to support both extremes of the mass-transfer efficiency. This is not unexpected, given that even in models that allow for RLA, mass transfer remains conservative thanks to efficient tidal coupling of the components. At longer orbital periods ($P_{\mathrm{orb}} \gtrsim 3\,$\,d), we start to see tension between the observations and the FCMT-based population, while the RLA-based population matches the observations fairly well.
At orbital periods of more than 30\,d, this becomes even clearer. Even though there are only three data points in this region of the diagram---W Crucis, RX Cassiopeiae, and SX Cassiopeiae---their mass ratios are lower than predicted by FCMT and agree quite well with the predictions of mass-transferring binaries with RLA.

As demonstrated in Appendix~F of \citetalias{Henneco2024a}, the mass-transfer efficiency of Case-A mass transfer with RLA for primary stars with initial masses in the range $\approx 5 \text{--} 20\,\msun$ is between ${\approx}30\text{--}60\%$, and for Case-B mass transfer this efficiency is ${\lesssim}25\%$. This implies again that we need a range of mass-transfer efficiencies, dictated by the binary configuration and evolutionary stage of the components, to explain the observed populations of Algols. 

We again note a system with an extreme mass ratio, namely $\upsilon$ Sagitarii. Unlike the systems HR~6819 and HR~2142 mentioned above, it is possible to reconcile this system with FCMT if we vary the primary mass within its $1\sigma$ uncertainty. 

\section{Discussion}\label{sec:discussion}

For FCMT, we find that, contrary to the RLA models in \citetalias{Henneco2024a}, the occurrence of contact is largely consistent with other model grids assuming FCMT, such as those in \citet{Pols1994}, \citet{deMink2007}, and \citet{Claeys2011}. For the RLA models in \citetalias{Henneco2024a}, only the outcomes of the initially closest Case-A models were consistent with those found in the aforementioned works. 

The main reason for the differences between the RLA and FCMT models is the increased number of initially wide-orbit systems that keep accreting matter owing to the absence of rotation-limited accretion. We reiterate that in the RLA models from \citetalias{Henneco2024a}, there exists large uncertainty on the fate of the Case-A and Case-B systems -- those that evolve into Algols and stripped-star binaries -- that avoid contact according to the \texttt{MESA} models. The combined luminosity of the binary components in the majority of these models was found to be insufficient to push the non-accreted matter away from the binary system to infinity (labelled `Non-conservative MT + cannot eject'). This was reconfirmed by \citet{Jin2026}. As argued in \citetalias{Henneco2024a} and \citet{Jin2026}, the interaction of the binary with this non-accreted and non-ejected matter could very well lead to enough orbital decay to induce a stellar merger.

By assuming fully conservative mass transfer in our models, we approach an upper limit for the occurrence of stellar mergers in mass-transferring binaries. Combined with the lower limit found from the RLA models in \citetalias{Henneco2024a}, we find that---depending on the mass-transfer efficiency---between $16\%$ and $38\%$ of binaries in our considered mass range at solar metallicity will likely merge. However, as mentioned in Sect.~\ref{sec:incidences}, this $38\%$ is likely still underestimating the merger fraction in this binary grid. As in the RLA grid, we only consider mergers occurring during the first contact phase. We also do not count possible mergers resulting from classical or double-core CE phases. Furthermore, there are still the $8\%$ of all models in this mass range for which we have no information about their fate due to numerical issues.

As described in Sect.~\ref{subsec:Initial10.2M}, the line between mass-transferring (Case-C) models and models not engaging in mass transfer has shifted slightly upward in the FCMT grid compared to the RLA grid due to the decreased resolution in $a_{\mathrm{i}}$ in this region of the initial parameter space. Because of this, the the relative increase in the total area of mass-transferring systems in $(q_{\mathrm{i}}\text{–}\log P_{\mathrm{i}})$-space is ${\sim}2\%$ (based on the shift of the line for the representative set of models with $M_{1,\,\mathrm{i}}=10.2\,\msun$) from the RLA grid to the  FCMT grid. Although this slight discrepancy is not ideal, it does not affect the comparison of incidences between the two grids, as done in Sect.~\ref{sec:incidences}, in a meaningful way.

The results in Sect.~\ref{sec:stable-mt} show a need for different mass-transfer efficiencies to explain different observed populations. This efficiency must be rooted in the physical state of the binary systems, and there are various mechanisms that may regulate the mass-transfer efficiency and are typically not considered in 1D stellar evolution models. One such mechanism is accretion via a disk with viscous stresses removing angular momentum to allow mass accretion beyond the rotational limit \citep{Paczynski1977, Paczynski1991, Popham1991}.
Moreover, outflows from such an accretion disk via jets and winds can further remove mass and angular momentum and thus explain the continued mass accretion in systems such as $\beta$ Lyrae \citep{Harmanec1996,Hoffmann1998,Umana2000,Ak2007}, W~Serpentis \citep{Piirola2005, Shepard2024} and even during star formation processes \citep{Popham1995, Popham1996, Bally2016}. Bipolar outflows stemming from disk instabilities could also explain the significant non-conservativeness required to explain the Algol population, as seen from radio observations of $\beta$ Lyr \citep{Umana2000}. Finally, as shown in \citet{Scherbak2025}, accretion disks can also have outflows from the outer Lagrange points. Such outflows can significantly influence subsequent mass transfer by removing matter with the specific angular momentum of the $\mathrm{L_{2/3}}$-point, modifying further orbital evolution of the binary system.

However, some Algols, including the prototype $\beta$~Persei, do not exhibit any evidence of an accretion disk, which might indicate that their non-conservativeness in mass transfer stems from rotationally-enhanced winds driving away excess mass past the critical rotation \citep{Packet1981} as used in many stellar evolution codes. Moreover, the interaction of the accretion stream with the disk can drive mass away, leading to subtle variations in the mass-transfer efficiency \citep{Armitage1998, Kunze2001}.

To get better insights into the intricacies of the mass transfer and accretion process, we need to resort to dedicated 3D simulations. Such simulations can capture the boundary-layer between the accretion disk and the accretor, mass and angular momentum ejection from the system, and the processes that govern the accretion rate onto the accretor. Finally, \citet{Picco2025} demonstrated that current binary evolution theory cannot explain systems such as HR6819, regardless of the mass transfer efficiency.

\section{Summary and conclusions}\label{sec:conclusions}

In this work, we present a grid of \texttt{MESA} binary evolution models similar to the grid described in \citet{Henneco2024a}. Contrary to the models in \citet{Henneco2024a}, in which accretion is limited by the spin-up of the accretor, the current models assume fully conservative mass transfer. Despite the simplifications in the applied method, \ie not modelling rotation and tides, the new grid allows us to explore the other extreme of the mass-transfer efficiency spectrum, where the models from \citet{Henneco2024a} cover the least efficient end. From a detailed comparison between the grids, we find that the continuous accretion in the fully-conservative mass transfer grid leads to the occurrence of contact phases all the way up to the initially widest binary systems, while most of these systems avoided contact when accretion was limited. Specifically, we find the occurrence of contact between objects of (super-)giant proportions in the current models with fully conservative mass transfer, which we labelled as double-core CEs. With fully conservative mass transfer, the incidence of systems that initiate contact is almost 6 times as high as with rotation-limited accretion (from $11\%$ to $62\%$) and the incidence of stellar mergers more than doubles (from $16\%$ to $38\%$). Whereas the merger incidence of $16\%$ from \citet{Henneco2024a} could only be interpreted as an absolute lower limit of the actual incidence of mergers in nature, we can now more confidently say that the real incidence is somewhere in the range of $16$--$38\%$, even with the upper limit of $38\%$ potentially still being underestimated.

From the two grids of binary evolution models, we create two synthetic binary populations, which we compare to observed Algol and stripped-star binaries. From this comparison, we found that, while stripped-star formation seems to favour more conservative mass transfer, this picture is less obvious for Algols that may require lower mass-transfer efficiencies. This reinforces the notion that not all stages of binary evolution have the same mass-transfer efficiency. Depending on the binary configurations, different physical mechanisms may regulate the efficiency of mass transfer; these mechanisms remain uncertain and partially unknown. Efforts are required to include new, physically motivated (\eg based on 3D simulation of mass transfer) prescriptions for accretion and accretor spin-up in 1D binary evolution codes that can self-consistently cover the variety of binary configurations.

\section*{Data availability}
The input files required to reproduce the models in this work and the machine-readable version of Table~\ref{tab:contact_tracing_results} are available at \href{https://zenodo.org/records/21804582}{https://zenodo.org/records/21804582} (\href{https://doi.org/10.5281/zenodo.21804582}{https://doi.org/10.5281/zenodo.21804582}). 

\begin{acknowledgements}
We thank the anonymous referee for their helpful and constructive feedback. We wish to thank Dandan Wei, Jim Fuller, Kareem El-Badry, Mike Lau, Pablo Marchant and Vincent Bronner for the helpful discussions and comments. Additional software used in this work includes \texttt{PyMesaReader} \citep{PyMesaReader2017}, \texttt{MPI for Python} \citep{Dalcin2005,Dalcin2021}, \texttt{Astropy} \citep{Astropy2013,Astropy2018,Astropy2022}, \texttt{NumPy} \citep{Harris2020}, and \texttt{SciPy} \citep{Scipy2020}. We used \texttt{Matplotlib} \citep{Hunter2007} for plotting. The authors acknowledge support from the Klaus Tschira Foundation. This work has received funding from the European Research Council (ERC) under the European Union’s Horizon 2020 research and innovation programme (Starting Grant agreement N$^\circ$ 945806: TEL-STARS). While funded by the European Union, views and opinions expressed are however those of the author(s) only and do not necessarily reflect those of the European Union or the European Research Council. Neither the European Union nor the granting authority can be held responsible for them. This work is supported by the Deutsche Forschungsgemeinschaft (DFG, German Research Foundation) under Germany’s Excellence Strategy EXC 2181/1-390900948 (the Heidelberg STRUCTURES Excellence Cluster). JH is grateful for support from UK Research and Innovation (UKRI) in the form of a Frontier Research grant under the UK government’s ERC Horizon Europe funding guarantee (SYMPHONY; PI Bowman, grant number: EP/Y031059/1). 
\end{acknowledgements}

\bibliographystyle{aa}
\bibliography{bibliography}

\begin{appendix}

\onecolumn

\section{Contact tracing results for other $M_{1,\,\mathrm{i}}$}\label{app:other_contact_tracing}
\begin{figure*}[h!]
\centering
  \includegraphics[width=16.2cm]{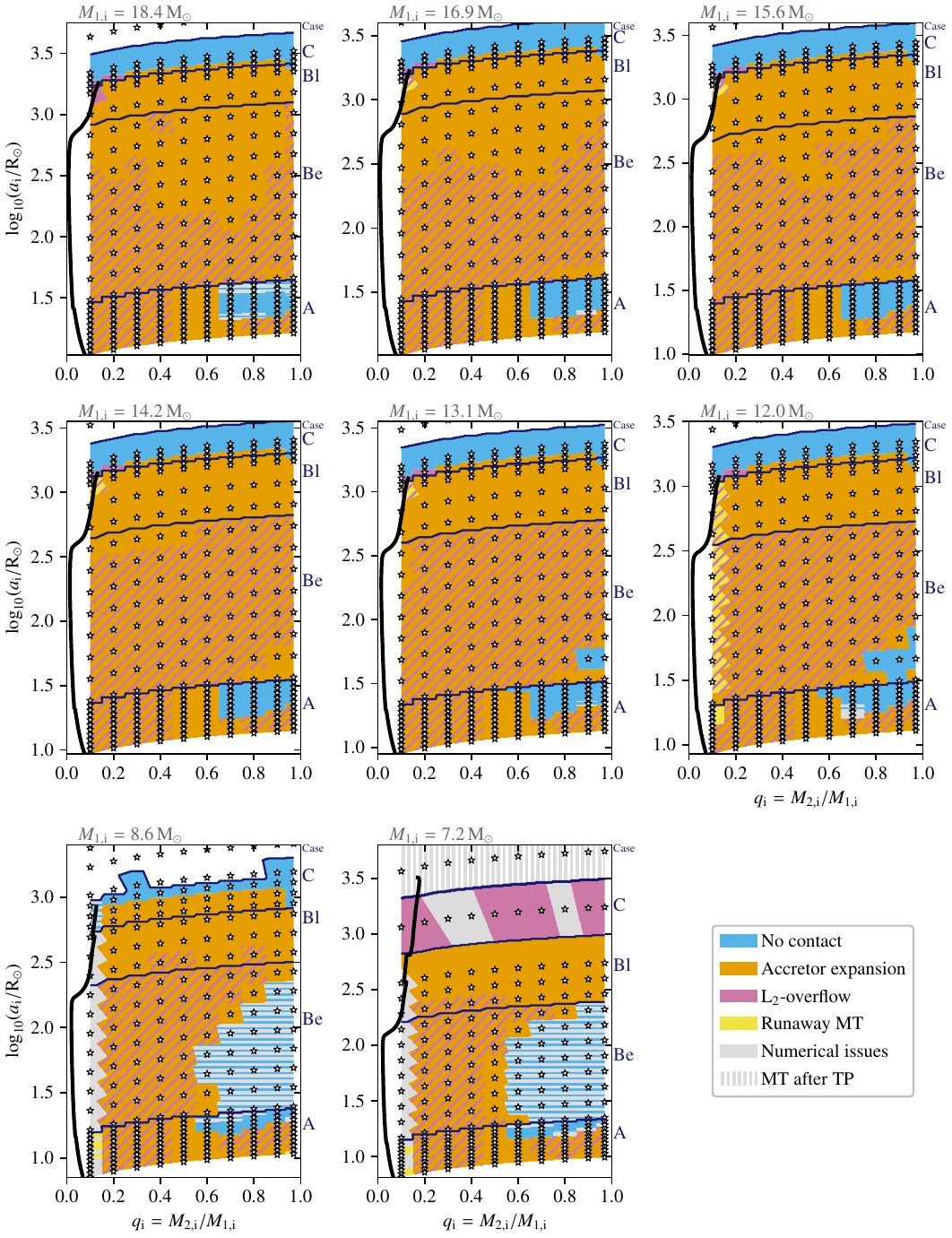}
    \caption{Contact tracing results for $M_{1,\,\mathrm{i}}=7.2\text{--}18.4\,\mathrm{M}_{\odot}$. Models with  $M_{1,\,\mathrm{i}}=7.2\,\mathrm{M}_{\odot}$ and initial separations larger than the Case-C systems experience numerical issues after the TP-AGB phase (see Sect.~\ref{sec:contact_tracing}). Only the model with $q_{\mathrm{i}} = 0.1$ avoids these issues and does not initiate mass transfer.}
     \label{fig:appendix_contact_tracing1}
\end{figure*}

\begin{figure*}
\centering
  \includegraphics[width=16.2cm]{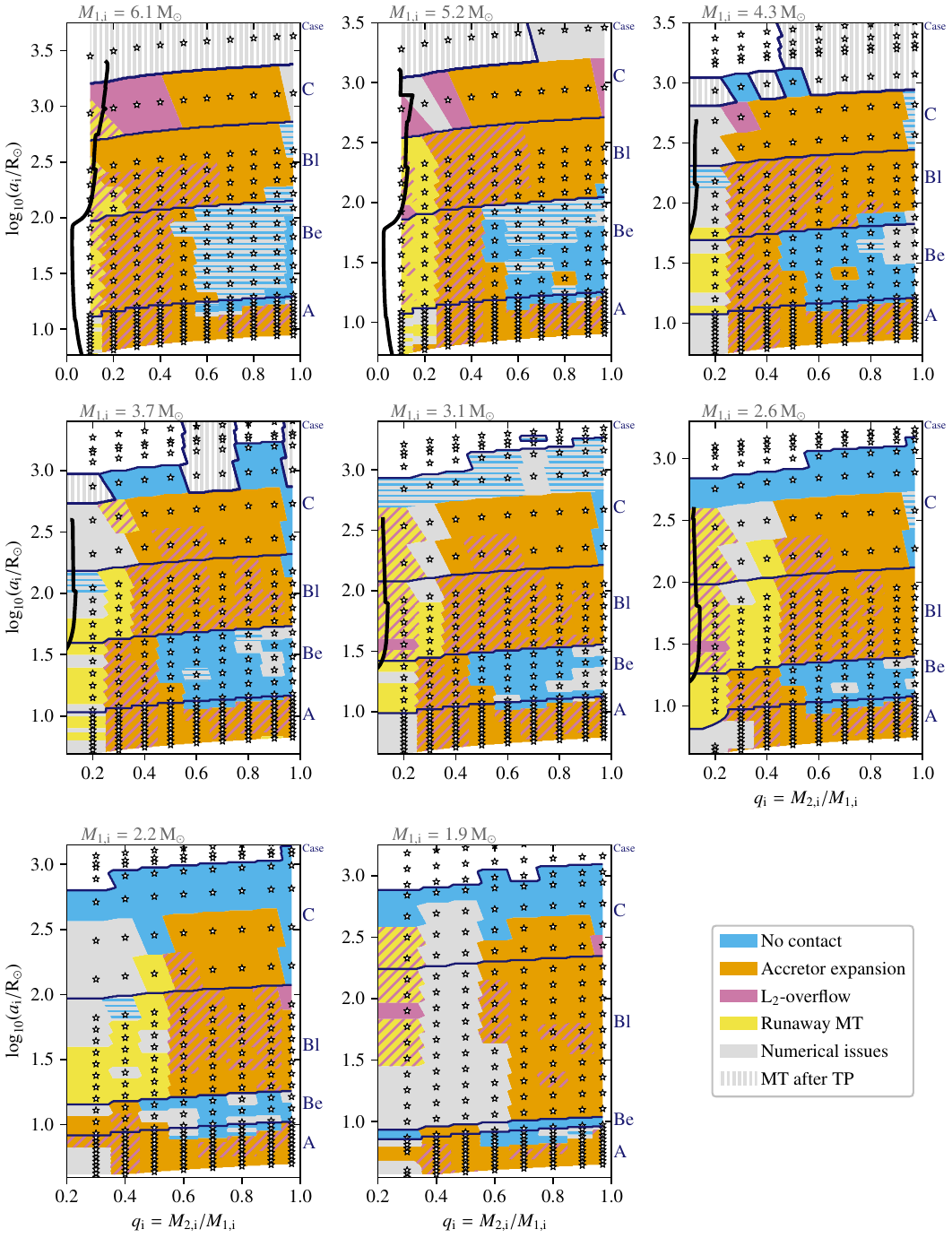}
    \caption{Contact tracing results for $M_{1,\,\mathrm{i}}=1.9\text{--}6.1\,\mathrm{M}_{\odot}$. Models with $M_{1,\,\mathrm{i}}=3.7\text{--}6.1\,\mathrm{M}_{\odot}$ experience numerical issues after the TP-AGB phase (see Sect.~\ref{sec:contact_tracing}). This explains the unexpected onset of mass transfer at initial separations larger than those of systems avoiding mass transfer. For $M_{1,\,\mathrm{i}}=1.9\text{--}2.2\,\mathrm{M}_{\odot}$ we find certain models where mass transfer starts when the primary is on the WD cooling track and experiences sudden radial expansion.}
     \label{fig:appendix_contact_tracing2}
\end{figure*}

\begin{figure*}
\centering
  \includegraphics[width=16.2cm]{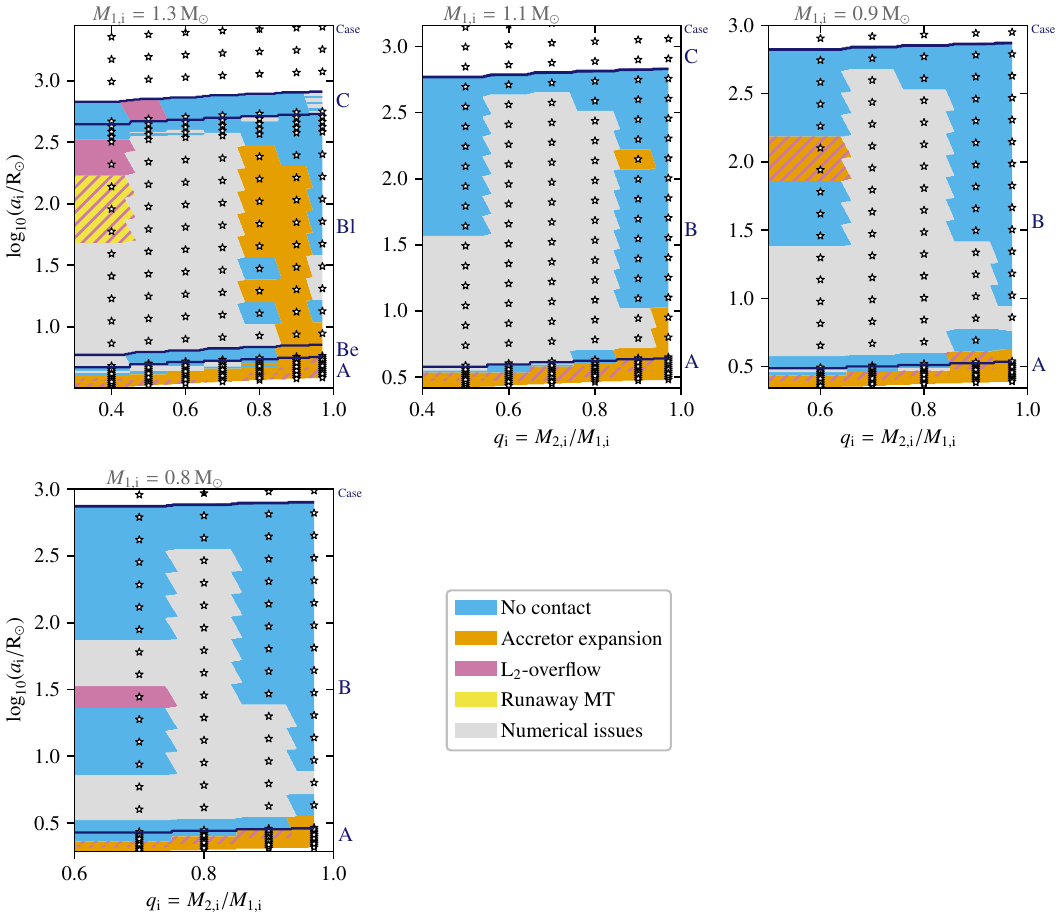}
    \caption{Contact tracing results for $M_{1,\,\mathrm{i}}=0.8\text{--}1.3\,\mathrm{M}_{\odot}$. For $M_{1,\,\mathrm{i}}=1.1\text{--}1.3\,\mathrm{M}_{\odot}$ we find certain models where mass transfer starts when the primary is on the WD cooling track and experiences sudden radial expansion.}
     \label{fig:appendix_contact_tracing3}
\end{figure*}

Figs.\,\ref{fig:appendix_contact_tracing1}--\ref{fig:appendix_contact_tracing3} contain the contact tracing results for the initial primary masses $M_{1,\,\mathrm{i}} = 0.8$, $0.9$, $1.1$, $1.3$, $1.9$, $2.2$, $2.6$, $3.1$, $3.7$, $4.3$, $5.2$, $6.1$, $7.2$, $8.6$, $12.0$, $13.1$, $14.2$, $15.6$, $16.9$, and $18.4\,\mathrm{M}_{\odot}$.

\FloatBarrier

\begin{sidewaystable*}

\section{Table with contact tracing results}\label{app:table}
Table \ref{tab:contact_tracing_results} contains an extract of the table containing the contact tracing results for all our models.

\caption{Extract of the table with the contact tracing results of all 5802 binary \texttt{MESA} models.}
\label{tab:contact_tracing_results}
\centering
\resizebox{\columnwidth}{!}{
\begin{tabular}{cccccccccccccccccc}
\hline
\begin{tabular}[c]{@{}c@{}}$M_{1,\,\mathrm{i}}$\tablefootmark{a}\\ $[\mathrm{M}_{\odot}]$\end{tabular} & \begin{tabular}[c]{@{}c@{}}$M_{2,\,\mathrm{i}}$\\ $[\mathrm{M}_{\odot}]$\end{tabular} & $\log_{10}(a_{\mathrm{i}}/\mathrm{R}_{\odot})$ & $\log_{10}(P_{\mathrm{i}}/\mathrm{d})$ & \begin{tabular}[c]{@{}c@{}}$M_{1,\,\mathrm{f}}$\tablefootmark{b}\\ $[\mathrm{M}_{\odot}]$\end{tabular} & \begin{tabular}[c]{@{}c@{}}$M_{2,\,\mathrm{f}}$\\ $[\mathrm{M}_{\odot}]$\end{tabular} & $\log_{10}(a_{\mathrm{f}}/\mathrm{R}_{\odot})$ & $\log_{10}(P_{\mathrm{f}}/\mathrm{d})$ & $\log_{10}(\mathrm{age}_{\mathrm{f}}/\mathrm{yrs})$ & AE\tablefootmark{c} & RMT\tablefootmark{d} & L2O\tablefootmark{e} & NC\tablefootmark{f} & MTTP\tablefootmark{g} & NI\tablefootmark{h} &  \begin{tabular}[c]{@{}c@{}}Case\\ {[}A,B,C{]}\end{tabular} & $\mathrm{ES}_{1}$\tablefootmark{i} & $\mathrm{ES}_{2}$\tablefootmark{j} \\ \hline
0.80 & 0.56 & 0.407 & -0.392 & 0.20 & 1.16 & 0.969 & 0.452 & 10.511 & 0 & 0 & 0 & 1 & 0 & 0 & [1,1,0] & post-MS & post-MS \\
0.95 & 0.57 & 2.745 & 3.090 & 0.53 & 0.57 & 2.885 & 3.372 & 10.479 & 0 & 0 & 0 & 1 & 0 & 0 & [0,1,0] & post-CHeB & MS \\          
1.12 & 0.90 & 0.616 & -0.165 & 0.21 & 1.80 & 1.453 & 1.092 & 9.945 & 0 & 0 & 0  & 1 & 0 & 0 & [1,1,0] & post-MS & post-MS \\     
1.33 & 0.53 & 2.629 & 2.872 & 0.53 & 0.53 & 2.874 & 3.363 & 9.632 & 0 & 0 & 0 & 1 & 0 & 0 & [0,1,1] & post-CHeB & MS \\          
1.58 & 1.11 & 0.681 & -0.130 & 1.11 & 1.58 & 0.680 & -0.131 & 9.430 & 1 & 0 & 0 & 0 & 0 & 0 & [1,1,0] & MS & MS \\          
1.87 & 1.68 & 2.254 & 2.169 & 1.75 & 1.79 & 2.253 & 2.168 & 9.142 & 1 & 0 & 0 & 0 & 0 & 0 & [0,1,0] & post-MS & MS \\          
2.21 & 0.66 & 1.554 & 1.166 & 2.20 & 0.67 & 1.550 & 1.160 & 8.927 & 0 & 1 & 0  & 0 & 0 & 0 & [0,1,0] & post-MS & MS \\          
2.62 & 1.83 & 0.934 & 0.140 & 1.37 & 3.08 & 1.046 & 0.308 & 8.860 & 0 & 0 & 0 & 0 & 0 & 1 & [1,0,0] & MS & post-MS \\     
3.10 & 1.24 & 0.832 & -0.007 & 3.07 & 1.27 & 0.818 & -0.028 & 8.376 & 1 & 0 & 1 & 0 & 0 & 0 & [1,0,0] & MS & MS \\
3.68 & 3.57 & 0.988 & 0.116 & 2.10 & 5.15 & 1.158 & 0.370 & 8.306 & 0 & 0 & 0 & 0 & 0 & 1 & [1,0,0] & MS & post-MS \\     
4.35 & 2.17 & 2.035 & 1.710 & 4.31 & 2.21 & 2.028 & 1.699 & 8.153 & 1 & 0 & 1 & 0 & 0 & 0 & [0,1,0] & post-MS & MS \\          
5.16 & 0.52 & 1.207 & 0.497 & 5.13 & 0.54 & 1.170 & 0.443 & 7.974 & 0 & 1 & 1 & 0 & 0 & 0 & [0,1,0] & post-MS & MS \\          
6.11 & 3.67 & 4.357 & 5.105 & 1.54 & 3.67 & 4.631 & 5.652 & 7.860 & 0 & 0 & 0 & 0 & 1 & 0 & [0,0,1] & post-CHeB & post-CHeB \\
7.24 & 5.79 & 1.202 & 0.309 & 2.67 & 10.18 & 1.568 & 0.861 & 7.746 & 0 & 0 & 0 & 0 & 0 & 1 & [1,0,0] & MS & post-MS \\     
8.57 & 5.14 & 2.385 & 2.072 & 7.06 & 6.57 & 2.335 & 1.999 & 7.507 & 1 & 0 & 1 & 0 & 0 & 0 & [0,1,0] & post-MS & MS \\          
10.16 & 6.10 & 1.399 & 0.557 & 9.11 & 6.96 & 1.368 & 0.512 & 7.372 & 1 & 0 & 0 & 0 & 0 & 0 & [0,1,1] & post-MS & MS \\        
12.03 & 10.83 & 1.25 & 0.253 & 5.520 & 17.00 & 1.51 & 0.659 & 7.292 & 0 & 0 & 0 & 0 & 0 & 1 & [1,0,0] & MS & post-MS \\     
13.14 & 6.57 & 2.637 & 2.371 & 12.46 & 6.97 & 2.619 & 2.348 & 7.187 & 1 & 0 & 1 & 0 & 0 & 0 & [0,1,0] & post-MS & MS \\          
14.25 & 11.40 & 1.243 & 0.224 & 6.45 & 18.77 & 1.478 & 0.580 & 7.211 & 0 & 0 & 0 & 0 & 0 & 1 & [1,0,0] & MS & post-MS \\     
15.57 & 15.10 & 1.403 & 0.425 & 3.36 & 24.85 & 2.225 & 1.676 & 7.173 & 0 & 0 & 0 & 1 & 0 & 0 & [1,1,1] & post-CHeB & MS \\         
16.88 & 1.69 & 3.262 & 3.323 & 14.93 & 1.69 & 3.311 & 3.419 & 7.074 & 0 & 0 & 0 & 1 & 0 & 0 & [0,0,1] & post-CHeB & MS \\          
18.44 & 14.75 & 2.152 & 1.532 & 15.90 & 16.65 & 2.153 & 1.537 & 6.989 & 1 & 0 & 1 & 0 & 0 & 0 & [0,1,0] & post-MS & MS \\          
20.0 & 19.40 & 3.510 & 3.532 & 14.99 & 16.63 & 3.606 & 3.723 & 6.986 & 0 & 0 & 0 & 1 & 0 & 0 & [0,0,1] & post-CHeB & CHeB \\       
... & ... & ... & ... & ... & ... & ... & ... & ... & ... & ... & ... & ... & ... & ... & ... & ... & ... \\
\hline
\end{tabular}
}
\tablefoottext{a}{`i' = `initial'}
\tablefoottext{b}{`f' = `final' -- at contact or termination}
\tablefoottext{c}{Accretor expansion}
\tablefoottext{d}{Runaway mass transfer}
\tablefoottext{e}{$\mathrm{L}_{2}$-overflow}
\tablefoottext{f}{No contact}
\tablefoottext{g}{Mass transfer after thermal pulses}
\tablefoottext{h}{Numerical issues}
\tablefoottext{i}{Primary's evolutionary state at contact or termination. MS: before core-H exhaustion. Post-MS: after core-H exhaustion and before core-He ignition. CHeB: after core-He ignition and before core-He exhaustion. Post-CHeB: after core-He exhaustion.}
\tablefoottext{j}{Secondary's evolutionary state at contact or termination.}
\end{sidewaystable*}

\end{appendix}

\end{document}